\documentclass[acmjetc,acmnow]{acmtrans2m}

\usepackage{times}

\usepackage{graphicx}
\usepackage{color}
\usepackage{dcolumn}
\usepackage{bm}
\usepackage{graphics}
\usepackage{lscape}
\usepackage{subfigure}

\def\permission{}

\title{ Arithmetic on a Distributed-Memory Quantum
Multicomputer\footnote{A shorter version of this paper appeared as
``Distributed Arithmetic on a Quantum Multicomputer'' in the
Proceedings of the International Symposium on Computer Architecture
(ISCA-33), 2006~\cite{van-meter05:_distr_arith_quant_multic}.}  }
\date{\today}

\author{RODNEY VAN METER\\
  Keio University and CREST-JST, Japan\\
\and W. J. MUNRO\\
  Hewlett-Packard Laboratories, Bristol, United Kingdom\\
\and KAE NEMOTO\\
  National Institute of Informatics, Tokyo, Japan\\
\and KOHEI M. ITOH\\
  Keio University and CREST-JST, Japan
}

\begin{abstract}
We evaluate the performance of quantum arithmetic algorithms run on a
distributed quantum computer (a quantum multicomputer).  We vary the
node capacity and I/O capabilities, and the network topology.  The
tradeoff of choosing between gates executed remotely, through
``teleported gates'' on entangled pairs of qubits (telegate), versus
exchanging the relevant qubits via quantum teleportation, then
executing the algorithm using local gates (teledata), is examined.  We
show that the teledata approach performs better, and that carry-ripple
adders perform well when the teleportation block is decomposed so that
the key quantum operations can be parallelized.  A node size of only a
few logical qubits performs adequately provided that the nodes have
two transceiver qubits.  A linear network topology performs acceptably
for a broad range of system sizes and performance parameters.  We
therefore recommend pursuing small, high-I/O bandwidth nodes and a
simple network.  Such a machine will run Shor's algorithm for
factoring large numbers efficiently.
\end{abstract}

\category{C.1.m}{Processor Architectures}{Miscellaneous}
\category{B.m}{Hardware -- MISCELLANEOUS}{}
            
\terms{Design} 
            
\keywords{Quantum computing, quantum computer architecture}
            
\begin{document}

\begin{bottomstuff} 
\end{bottomstuff}
            
\maketitle

\section{Introduction}

We are investigating the design of a {\em quantum multicomputer}, a
machine consisting of many small quantum computers connected together
to cooperatively solve a single
problem~\cite{van-meter05:_distr_arith_quant_multic,van-meter06:thesis}.
Such a system may overcome the limited capacity of quantum computing
technologies expected to be available in the near term, scaling to
levels which dramatically outperform classical computers on some
problems~\cite{nielsen-chuang:qci,shor:factor,grover96,deutsch-jozsa92}.

The main question in considering a multicomputer is whether the
system performance will be acceptable {\em if} the implementation
problems can be solved.  We focus on distributed implementation of
three types of arithmetic circuits derived from known classical adder
circuits~\cite{vedral:quant-arith,cuccaro04:new-quant-ripple,draper04:quant-carry-lookahead,ercegovac-lang:dig-arith}.
For many algorithms, notably Shor's algorithm for factoring large
numbers, arithmetic is an important component, and integer addition is
at its core~\cite{shor:factor,van-meter04:fast-modexp}.  Our
evaluation criterion is the latency to complete the addition.  The
goal is to achieve ``reasonable'' performance for Shor's factoring
algorithm for numbers up to a thousand bits.

Our distributed quantum computer must create a shared quantum state
between the separate nodes of our machine. As we perform our
computation, this quantum state evolves and we are dependent on either
quantum teleportation of data qubits or teleportation-based remote
execution of quantum gates to create that shared
state~\cite{bennett:teleportation,gottsman99:universal_teleport}.

The nodes of the machine may be connected in a variety of topologies,
which will influence the efficiency of the algorithm.  We concentrate
on only three topologies (shared bus, line, and fully connected) and
two additional variants (2bus, 2fully), constraining our engineering
design space and deferring more complex topology analysis for future
work.  Our analysis is done attempting to minimize the required number
of qubits in a node while retaining reasonable performance; we
investigate node sizes of one to five logical qubits per
node.

In this research we show that:
\begin{itemize}
\item teleportation of data is better than teleportation of gates;
\item decomposition of teleportation brings big benefits in
  performance, making a carry-ripple adder effective even for large
  problems;
\item a linear topology is an adequate network for the foreseeable
  future; and
\item small nodes (only a few logical qubits) perform acceptably, but
  I/O bandwidth is critical.
\end{itemize}
A multicomputer built around these principles and based on solid-state
qubit technology will perform well on Shor's algorithm.  These results
collectively represent a large step in the design and performance
analysis of distributed quantum computation.

We begin at the foundations, including related work and definitions of
some of the terms we have used in this introduction.  Next, we discuss
our node and interconnect architectures, followed by mapping the
arithmetic algorithms to our system.  Performance estimates are
progressively refined, including showing how decomposing the
teleportation operation makes the performance of the CDKM carry-ripple
adder competitive with the carry-lookahead adder.  We conclude with
specific recommendations for a medium-term goal of a modest-size
quantum multicomputer.

\section{Foundations}
\label{sec:foundations}

A quantum computer is a machine that uses quantum mechanical effects
to achieve potentially large reductions in the computational
complexity of certain
tasks~\cite{nielsen-chuang:qci,shor:factor,grover96,deutsch-jozsa92}.
Quantum computers exist, but are slow, very small (consisting of only
a few quantum bits, or \emph{qubits}) and not reliable.  Also, they
have very limited
scalability~\cite{vandersypen:shor-experiment,gulde03:_implem_deuts_jozsa}.
True architectural research for a large-scale quantum computer can be
said to have only just
begun~\cite{van-meter:qarch-impli,oskin:quantum-wires,copsey:q-com-cost,kielpinski:large-scale,steane04:_gop-qc,kim05:_system,balensiefer:isca05}.

Classically, the best known algorithm for factoring large numbers is
$O(e^{(nk \log^2 n)^{1/3}})$, where $n$ is the length of the number,
in bits, and $k = (\frac{64}{9} + \epsilon)\log 2$, whereas Shor's
quantum factoring algorithm is polynomial ($O(n^3)$ or
better)~\cite{knuth:v2-3rd,shor:factor,van-meter04:fast-modexp}.
These gains are achieved by taking advantage of {\em superposition} (a
quantum system being in a complex linear combination of states, rather
than the single state that is possible classically), {\em
entanglement} (loosely speaking, the state of two quanta not being
independent), and {\em interference} of the quantum wave functions
(analogous to interference in classical wave mechanics).  Of these,
only entanglement of pairs of qubits, as the core of quantum
teleportation, is directly relevant to this paper.  Otherwise, only a
limited familiarity with quantum computing is required to understand
this paper, and we introduce the necessary terminology and background
in this section.  Readers interested in more depth are referred to
popular~\cite{williams:_ultim_zero_one} and
technical~\cite{nielsen-chuang:qci} texts on the subject.

Teleportation of quantum states (qubits, or quantum data) has been
known for more than a decade~\cite{bennett:teleportation}.  It has
been demonstrated
experimentally~\cite{furusawa98,bouwmeester:exp-teleport}, and has
been suggested as being necessary for moving data long distances
within a single quantum computer~\cite{oskin:quantum-wires}.
Teleportation consumes Einstein-Podolsky-Rosen pairs, or {\em EPR
pairs}.  EPR pairs are pairs of particles or qubits which are {\em
entangled} so that actions on one affect the state of the other.  EPR
pairs can be created in a variety of ways, including reactions that
simultaneously emit pairs of photons whose characteristics are
related and many quantum gates on two qubits.  Entanglement is a
continuous, not discrete, phenomenon, and several weakly entangled
pairs can be used to make one strongly entangled pair using a process
known as {\em
  purification}~\cite{cirac97:_distr_quant_comput_noisy_chann}.

\subsection{Qubus Entanglement Protocol}

\begin{figure}
\centerline{\hbox{
\includegraphics[width=8cm]{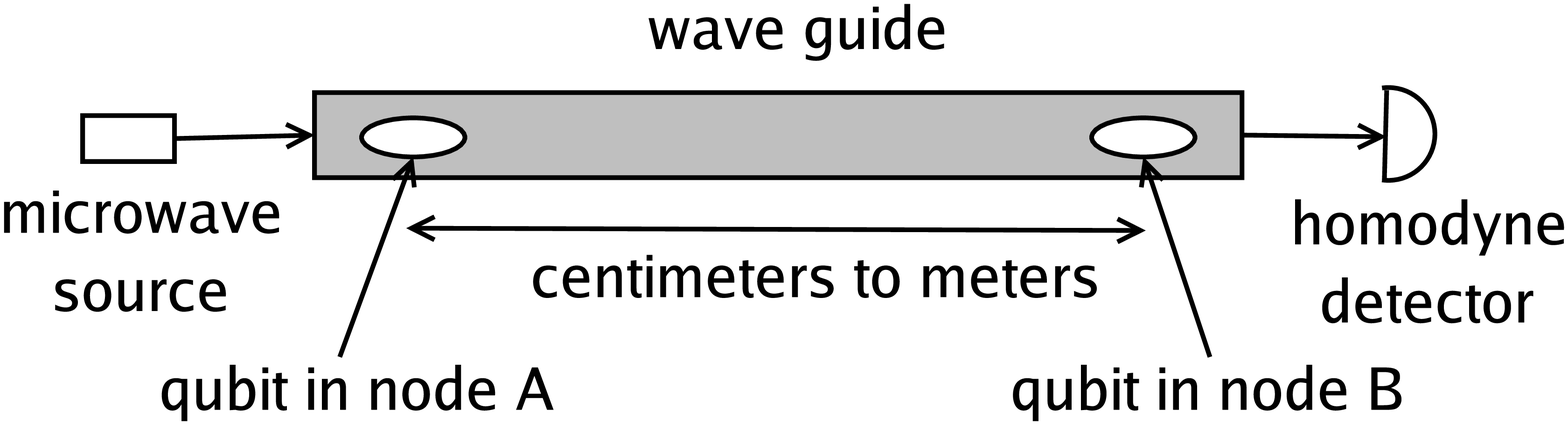}}}
\caption{Qubus link block diagram.}
\label{fig:qubus}
\end{figure}

Our approach to creating EPR pairs contains no direct qubit-qubit
interactions and does not require the use of single photons, instead
using laser or microwave pulses as a probe
beam~\cite{nemoto04:_nearly_deter,munro05:_weak}.  Two qubits are
entangled indirectly through the interaction of qubits with a common
quantum field mode created by the probe beam -- a continuous quantum
variable -- which can be thought of as a communication bus, or
``qubus''~\cite{spiller05:_qubus}.  We call the qubus-qubit
entanglement protocol {\em QEP}.  A block diagram of a qubus link is
shown in Figure~\ref{fig:qubus}.  For our purposes in the quantum
multicomputer, the qubits are likely to be separated by centimeters to
meters, though the protocol is expected to work at the micron scale
(within a chip)~\cite{spiller05:_qubus} and at the WAN scale
(kilometers)~\cite{van-loock06:_hybrid_quant_repeater}.

For some solid state qubit systems, the interaction with a bus mode
takes the effective form of a cross-Kerr nonlinearity, analogous to
that for optical systems, described by an interaction Hamiltonian of
the form
\begin{equation}
\label{Hintck}
H_{int} = \hbar \chi \sigma_z a^{\dagger} a.
\end{equation}
When acting for a time $t$ on a qubit-bus system, this interaction
effects a rotation (in phase space) by an angle $\pm \theta$ on a qubus
coherent state, where $\theta = \chi t$ and the sign depends on the
qubit computational basis amplitude.  By interacting the probe beam
with the qubit, the probe beam picks up a $\theta$ phase shift if it
is in one basis state (e.g., $|0\rangle$) and a $-\theta$ phase shift
if it is in the other (e.g., $|1\rangle$).  If the same probe beam
interacts with two qubits, it is straighforward to see that the probe
beam acting on the two-qubit states $|0\rangle |1\rangle$ and
$|1\rangle |0\rangle$ picks up no net phase shift because the
opposite-sign shifts cancel, while the probe beam acting on the states
$|0\rangle |0\rangle$ and $|1\rangle |1\rangle$ picks up phase shift
$\pm 2\theta$. An appropiate measurement determines whether the probe
beam has been phase shifted (in effect taking the absolute value of
the shift), projecting the qubits into either an even parity state or
an odd parity state.  The measurement shows only the parity of the
qubits, not the actual values, leaving them in an entangled state.
This state can be then used as our EPR pair.

\subsection{Teleporting Gates and Teleporting Data}
\label{sec:telethisnthat}

\begin{figure}
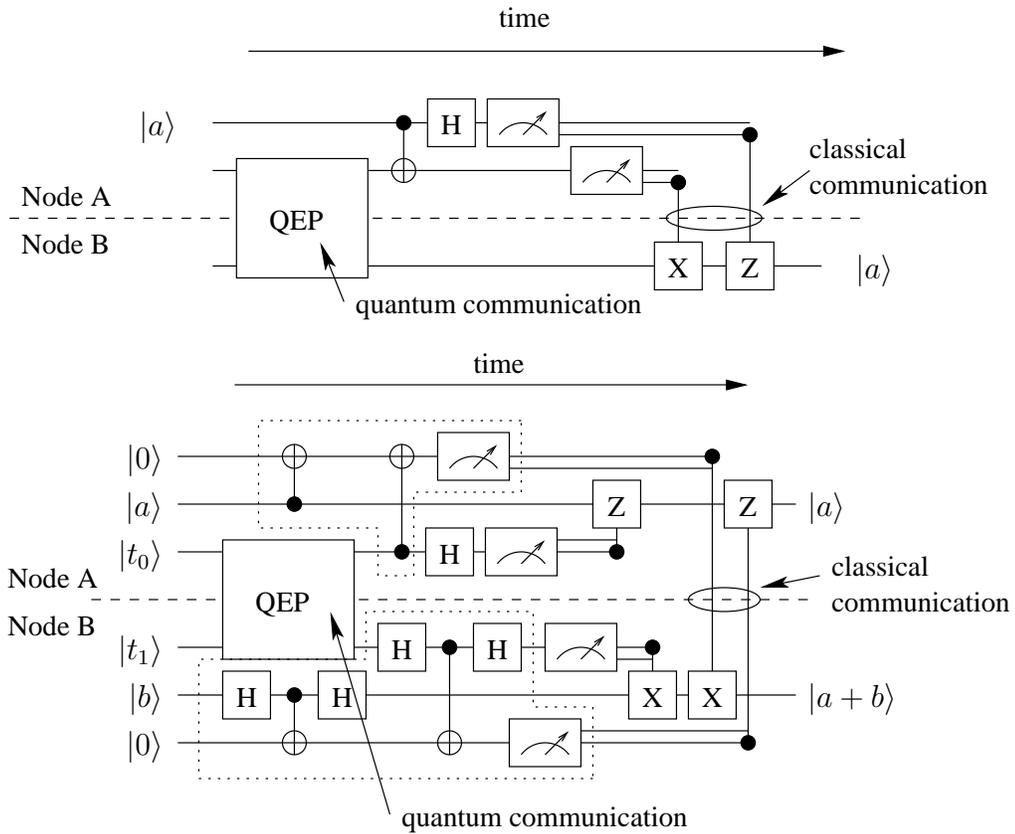

\centering
\subfigure{
\input{basic-teleportation.pstex_t}}
\vspace{.1in}
\subfigure{
\input{qep-telegate-corrected.pstex_t}}
\caption{A teleportation circuit (top) and teleported control-\textsc{not}
  (\textsc{cnot}) gate (bottom).  Time flows left to right, each horizontal
  line represents a qubit, and each vertical line segment with
  terminals is a quantum gate.  A segment with a $\oplus$ terminal is
  a control-\textsc{not} (\textsc{cnot}).  The ``meter'' box is measurement of a qubit's
  state.  The boxes with H, X, and Z in them are various qubit
  gates. The large box labeled QEP is the qubus EPR pair
  generator. (See Nielsen and Chuang for more details on the
  notation.)}
\label{fig:telebasics}
\end{figure}

\begin{figure}
\centerline{\hbox{
\includegraphics[width=8cm]{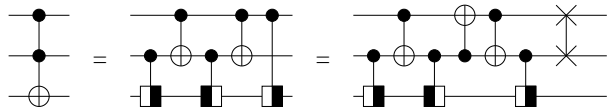}}}
\caption{\textsc{ccnot} (control-control-\textsc{not}, or Toffoli) gate
  constructions for our architectures.The leftmost object is the
  canonical representation of this three-qubit gate.  The rightmost
  construction we use for the line topology; the middle construction
  we use for all other topologies.  The box with the bar on the right
  represents the square root of $X$ (\textsc{not}), and the box with the bar on
  the left its adjoint.  The last gate in the rightmost construction
  is a SWAP gate, which exchanges the state of two qubits.}
\label{fig:5gcc}
\end{figure}

To teleport a qubit, one member of the EPR pair is held locally, and
the other by the teleportation receiver.  The qubit to be teleported
is entangled with the local EPR member, then both of those are {\em
measured}, which will return 0 or 1 for each qubit.  The results of
this measurement are transmitted to the receiver, which then executes
gates locally on its member of the EPR pair, conditional on the
measurement results, recreating the (now destroyed) original state at
the destination.  The circuit for teleportation is shown in
figure~\ref{fig:telebasics}.

Gottesman and Chuang showed that teleportation can be used to
construct a control-\textsc{not} (\textsc{cnot})
gate~\cite{gottsman99:universal_teleport}.  Their original teleported
gate requires two EPR pairs.  We use an approach based on parity gates
that consumes only one EPR pair, as shown in
figure~\ref{fig:telebasics}~\cite{munro05:_weak}.  Locally, the parity
gates can be implemented with two \textsc{cnot} gates and a measurement
(outlined with dotted lines in the figure).  Double lines are
classical values that are the output of the measurements; when used as
a control line, we decide classically whether or not to execute the
quantum gate, based on the measurement value.  The last gate involves
classical communication of the measurement result between nodes.  As
shown, this construction is not fault tolerant; it must be built over
fault-tolerant gates.  Alternatively, the qubus approach can be used
as the node-internal interconnect.  Its natural gate is the parity
gate, and is fault tolerant.

In designing algorithms for our quantum multicomputer, therefore, we
have a choice: when two qubits in different nodes of our multicomputer
are required to interact, we can either move data (qubits) from one
node to another, then perform the shared gate, or we can use a
teleported gate directly on the qubits, without moving them.  We will
call the data-moving approach {\em teledata} and the
teleportation-based gate approach {\em telegate}.

For some algorithms, we can use a simple, visual approach to counting
the number of remote operations necessary to execute the algorithm
using either the teledata or telegate approach (see
section~\ref{sec:algorithm}).  For the telegate approach, we assign a
cost of three to each two-node Toffoli (control-control-\textsc{not})
gate, and each three-node Toffoli gate we count as five.  The
three-node Toffoli gate should cost more, as in figure~\ref{fig:5gcc},
but pipelining of operations across multiple nodes hides the
additional latency.  We assign two-node \textsc{cnot} gates a cost of
one.\footnote{There are other possible decompositions of the Toffoli
gate, but the differences are less than a factor of two.  Which
approach is best will depend on the choice of quantum error correction
(QEC), as some are more difficult to implement on encoded
qubits~\cite{divincenzo98:_gates-and-circuits}.}

\subsection{Distributed Quantum Computation}

Early suggestions of distributed quantum computation include
Grover~\cite{grover97:_quant_telec}, Cirac et
al.~\cite{cirac97:_distr_quant_comput_noisy_chann}, and Steane and
Lucas~\cite{steane:ion-atom-light}.  A recent paper has proposed
combining the cluster state model with distributed
computation~\cite{lim05:_repeat_until_success}.  Such a distributed
system generally requires the capability of transferring qubit state
from one physical representation to another, such as nuclear spin
$\leftrightarrow$ electron spin $\leftrightarrow$
photon~\cite{mehring03:_entan-electron-nuke,jelezko04:_observ,childress05:_ft-quant-repeater}.

Yepez distinguished between distributed computation using entanglement
between nodes, which he called type I, and without inter-node
entanglement (i.e., classical communication only), which he called
type II~\cite{yepez01:_type_ii}.  Our quantum multicomputer is a type
I quantum computer.  Jozsa and Linden showed that Shor's algorithm
requires entanglement across the full set of qubits, so a type II
quantum computer cannot achieve exponential
speedup~\cite{jozsa03:entangle-speedup,love05:_type_ii_quant_algo}.
Much of the work on our multicomputer involves creation and management
of that shared entanglement.

Yimsiriwattana and Lomonaco have discussed a distributed version of
Shor's algorithm \cite{yimsiriwattana04:dist-shor}, based on one form
of the Beckman-Chari-Devabhaktuni-Preskill modular exponentiation
algorithm~\cite{beckman96:eff-net-quant-fact}.  The form they use
depends on complex individual gates, with many control variables,
inducing a large performance penalty compared to using only two- and
three-qubit gates.  Their approach is similar to our telegate
(sec.~\ref{sec:telethisnthat}), which we show to be slower than
teledata.  They do not consider differences in network topology, and
analyze only circuit size, not depth (time performance).

\section{Node and Interconnect Architecture}

A multicomputer~\cite{athas:multicomputer} is a constrained form of
distributed system.  All parts of the system are geographically
colocated.  Short travel distances (up to a few tens of meters)
between nodes reduce latency, simplify coordinated control of the
system, and increase signal fidelity.  We assume a regular network
topology, a dedicated network environment, and scalability to
thousands of nodes.  We concentrate on a homogeneous node technology
based on solid-state qubits, with a qubus interconnect, though our
results apply to essentially any choice of node and interconnect
technologies, such as single photon-based qubit
transfer~\cite{wallraff04:_strong-coupling,matsukevich:matter-light-xfer}.

Future, larger quantum computers will be built on technologies that
are inherently limited in the number of qubits that can be
incorporated into a single
device~\cite{nielsen-chuang:qci,spiller:qip-intro-cp,van-meter:qarch-impli,arda:qc-roadmap-v2}.
The causes of these limitations vary with the specific technology, and
in most cases are poorly understood, but may range from the low tens
to perhaps thousands; integration of the densities we are accustomed
to in the classical world is not even being seriously discussed for
most technologies.  For example, flux Josephson junction qubits, which
are built using VLSI chip manufacturing technology, may be 100 microns
square; even a large chip would hold only a few thousand physical
qubits~\cite{martinis02:_rabi_oscil_phase}.  The scalable ion trap,
built with a PC board fabrication process, requires an even larger
area for all of the control structures to manage each trapped
atom~\cite{kielpinski:large-scale}.  Quantum error correction (QEC)
naturally reduces the number of available logical (application-level)
qubits by a large
factor~\cite{shor:qecc,calderbank96:_good-qec-exists,steane02:ft-qec-overhead}.
Two levels of the Steane 7-qubit code, for example, which encodes a
single logical qubit in seven lower-layer qubits, would impose a 49:1
encoding and storage penalty.  Even aggressive management of the
overhead imposed by error correction may still leave an ion trap
system with a surface area of a large fraction of a square
meter~\cite{thaker06:_cqla}.  Such a system would be difficult to
fabricate and operate.  Therefore, it makes sense to examine the
utility of a device that can hold only a few logical qubits,
especially if the device can create shared entanglement with another
similar device.

We choose a node technology based on solid-state qubits, such as
Josephson-junction superconducting
qubits~\cite{nakamura99:_coher_cooper,wallraff04:_strong-coupling,johansson05:_vacuum_rabi_lc}
or quantum dots~\cite{fujisawa98:double-dot}, which will require a
microwave qubus.  Each node has many qubits which are private to the
node, and a few transceiver qubits that can communicate with the
outside world.  Node size is limited by the number of elements that
can practically be built into a single device, including control
structures, external signalling, packaging, cooling, and shielding
constraints.  The primary advantages of these solid-state technologies
are their speed, with physical gate times in the low nanoseconds, and
their potential physical scalability based on photolithographic
techniques.

Hollenberg's group has recently proposed a multicomputer using small
ion traps as the nodes~\cite{oi06:_dist-ion-trap-qec}.  Each node will
contain enough physical qubits to hold one logical qubit plus a few
ancillae and a transceiver qubit, corresponding to our ``baseline''
case, discussed below.  A single photon-based entanglement scheme will
be used.  They do not explore algorithms, concentrating on the error
correcting parts of the system, and do not discuss the details of the
``optical multiplexer'' in their system that corresponds to our
interconnect.  This system has the advantage that it could be
implemented using existing technology.

Throughout this paper, qubits and operations on them are understood to
be logical.  Although the QEP protocol in theory supports EPR pair
creation over many kilometers, our design goal is a scalable quantum
computer in one location (such as a single lab).  We consider a 10nsec
classical communication latency, corresponding roughly to 2m distance
between nodes.  We find that performance is insensitive to this
number.

\begin{figure}
\centerline{\hbox{
\includegraphics[width=12cm]{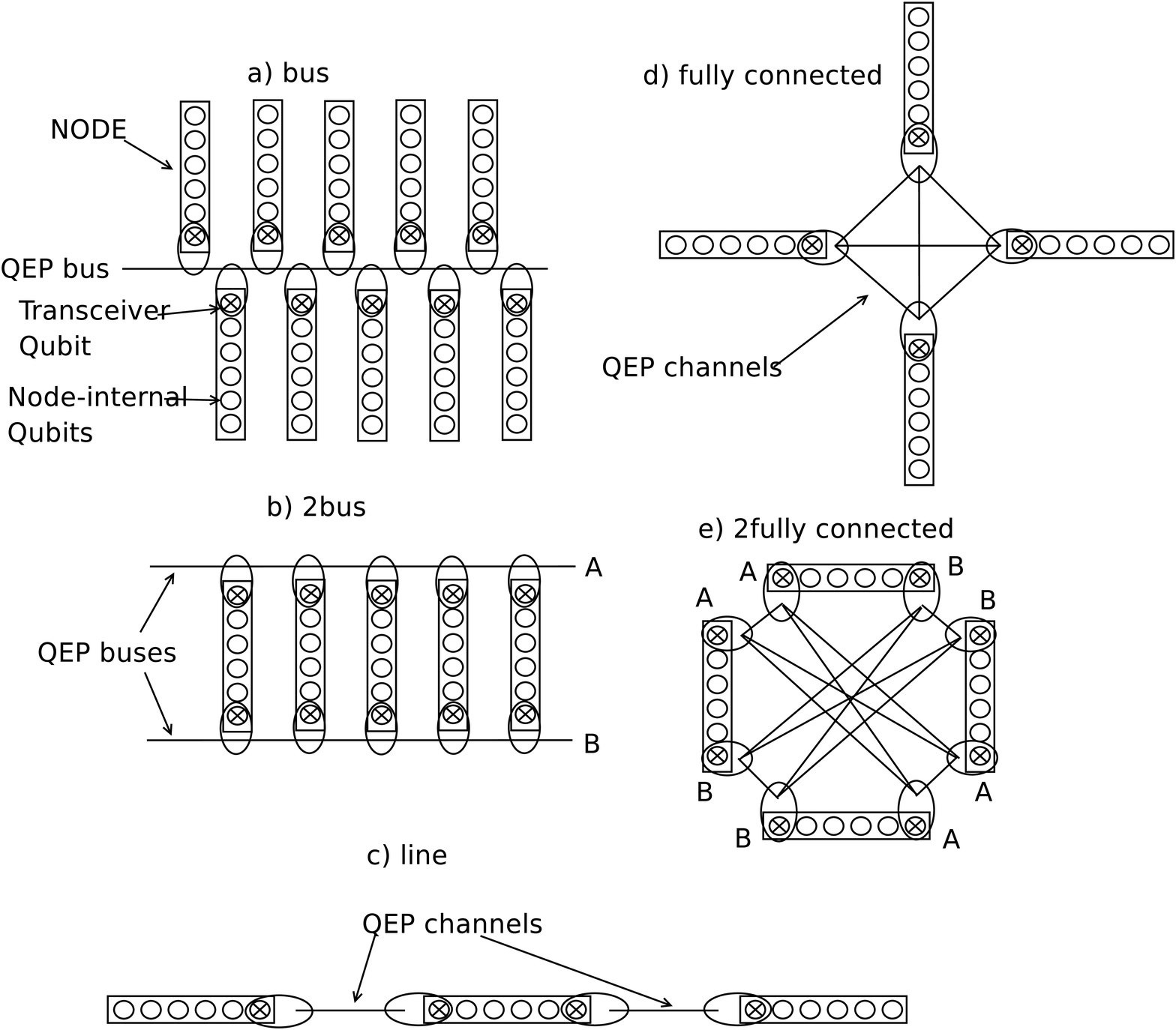}}}
\caption{The five physical topologies analyzed in this paper.}
\label{fig:five-topos}
\end{figure}

We consider five interconnect networks: shared bus, line of nodes,
fully connected, two-transceiver bus (2bus), and two-transceiver fully
connected (2fully) as in figure~\ref{fig:five-topos}.  For the shared
bus, all nodes are connected to a single bus.  Any two nodes may use
the bus to communicate, but it supports only a single transaction at a
time.  For 2bus, each node contains two transceiver qubits and
connects to two independent buses, labeled ``A'' and ``B'' in the
figure, that may operate concurrently.  In the line topology, each
node uses two transceiver qubits, one to connect to its left-hand
neighbor and one to connect to its right-hand neighbor.  Each link
operates independently, and all links can be utilized at the same
time, depending on the algorithm.  For the fully-connected network,
each node has a single transceiver qubit which can connect to any
other node without penalty via some form of classical circuit-switched
network, though of course each transceiver qubit can be involved in
only one transaction at a time.  Although we have characterized this
network as ``fully connected'', implying an $n\times n$ crossbar
switch, any non-blocking network, such as a Clos network, will do,
provided that the signal loss through each stage of the network is not
significant~\cite{dally04:_interconnects}.  The network can also be
optimized to match the particular traffic pattern, though that is
unlikely to be necessary.  The 2fully topology utilizes two
transceiver qubits per node for concurrent transfers on two separate
networks, one connecting the transceivers labeled ``A'' and one
connected the transceivers labeled ``B''.  Many mappings of qubits to
nodes and gates to bus timeslots are possible; we do not claim the
arrangements presented here are optimal.

The effective topology may be different from the physical topology,
depending on the details of a bus transaction.  For example, even if
the physical topology is a bus, the system may behave as if it is
fully connected if the actions {\em internal} to a node to complete a
bus transaction are much longer than the activities on the bus itself,
allowing the bus to be reallocated quickly to another transaction.
Some technologies may support frequency division multiplexing on the
bus, allowing multiple concurrent transactions.

This research is part of an overall effort to design a scalable
quantum multicomputer.  Elsewhere, we have investigated distributed
quantum error correction, determining that two layers of the Steane
[[7,1,3]] quantum error correction code (for a total capacity penalty
of 49:1) will protect against error rates up to $\sim 1\%$ in the
teleportation process.  We have also found that each link in the
interconnect may be serial, causing only a small penalty in
performance and reliability, while substantially simplifying the
hardware~\cite{van-meter07:_commun_links_distr_quant_comput,van-meter06:thesis}.
Those results help constrain the hardware of the quantum
multicomputer; in this paper, we analyze the software and
performance.

\section{Algorithm}
\label{sec:algorithm}

The introduction of Shor's factoring algorithm spurred interest in
arithmetic circuits for quantum computers.  Several groups almost
immediately began investigating the modular exponentiation phase of
the
algorithm~\cite{vedral:quant-arith,beckman96:eff-net-quant-fact,miquel1996fdq}.
All of these early algorithms used various types of carry-ripple
adders, which are $O(n)$ in both depth and gate complexity, and
composed modular multiplication and exponentiation in the most
straightforward fashion.  Shortly thereafter, other types of adders
and various other optimizations were
introduced~\cite{gossett98:q-carry-save,draper00:quant-addition,cleve:qft,beauregard03:small-shor,zalka98:_fast_shor}.
Some of these circuits operate on Fourier-transformed
numbers~\cite{beauregard03:small-shor}; while these circuits use fewer
qubits than the original Vedral-Barenco-Ekert (VBE) style of
carry-ripple adder~\cite{vedral:quant-arith}, implementing the small
rotations that are necessary for the Fourier transform may be
difficult on QEC-encoded qubits.  Zalka examined the
Sch\"onhage-Strassen FFT-based multiplication algorithm and found it to
be faster than the ``obvious'' approach only for factoring numbers
larger than 8 kilobits~\cite{zalka98:_fast_shor}; our own analysis
places the crossover at closer to 32kb.  Some of these approaches are
evaluated in more detail in our previous
work~\cite{van-meter04:fast-modexp}.  In this paper, we choose to
concentrate on integer addition only, as the fundamental building
block of arithmetic.

Recently, several new addition circuits have been introduced, some
based on standard classical
techniques~\cite{ercegovac-lang:dig-arith}.  The
Cuccaro-Draper-Kutin-Moulton (CDKM) carry-ripple
adder~\cite{cuccaro04:new-quant-ripple} is faster than the original
VBE adder and uses fewer qubits.  The advantage of the Draper
Fourier-based adder was its use of fewer qubits; the development of
CDKM makes the Fourier adder less attractive due to its complex
implementation.  Takahashi and Kunihiro, working from CDKM, have
eliminated the need for ancillae, at the expense of a much deeper (but
still $O(n)$) circuit~\cite{takahashi05:adder}.  The carry-lookahead
adder~\cite{draper04:quant-carry-lookahead} is $O(\log n)$ depth and
$O(n)$ gate complexity.  Our own carry-select and conditional-sum
adders~\cite{van-meter04:fast-modexp} are $O(\sqrt{n})$ and $O(\log
n)$ circuit depth, respectively, but use more ancillae.  The
definitions of these algorithms ignore communications costs; in most
real systems, distant qubits cannot interact directly, and this
impacts performance.  The carry-lookahead adder and Takahashi-Kunihiro
adder do not present obvious mappings onto architectures with such
limitations.

From among these, we have chosen to evaluate three different addition
algorithms: VBE~\cite{vedral:quant-arith}, the
Cuccaro-Draper-Kutin-Moulton (CDKM) carry-ripple
adder~\cite{cuccaro04:new-quant-ripple}, and the carry-lookahead
adder~\cite{draper04:quant-carry-lookahead}.  In this section we
discuss the adders without regard to the network topology; the
following section presents numeric values for different topologies and
gate timings.  None of these circuits has been optimized for a system
in which accessing some qubits is very fast and accessing others is
very slow, as in our multicomputer; it is certainly possible that
faster circuits for our proposed system will be found.

\subsection{Carry-Ripple Adders}

\begin{figure*}
\input{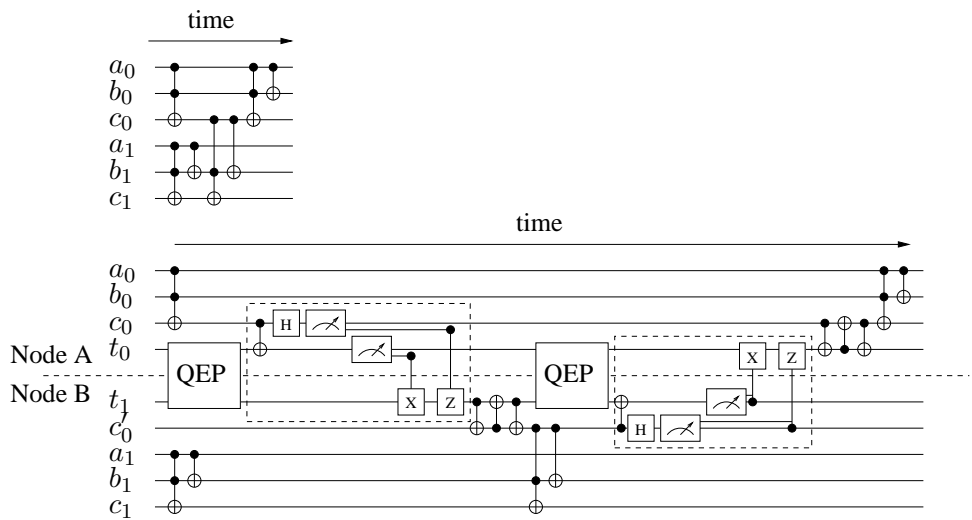}
\caption{Details of a distributed 2-qubit VBE adder.  The top circuit
is the monolithic form, and the bottom circuit is the distributed form
using the teledata method.  The solid box (QEP) is the qubus EPR pair
generator; the circuits in dashed boxes are standard quantum
teleportation circuits.  The ``meter'' box is measurement of a qubit's
state.  The boxes with H, X, and Z in them are various single-qubit
gates.}
\label{fig:teleport-2bit-adder}
\end{figure*}

Figure~\ref{fig:teleport-2bit-adder} shows a two-qubit VBE
carry-ripple adder~\cite{vedral:quant-arith} in its monolithic (top)
and distributed (bottom) forms.  Each horizontal line represents a
qubit; the \emph{ket} notation is omitted for clarity.  The QEP block
creates an EPR pair.  The dashed boxes delineate the teleportation
circuit, which is assumed to be perfect.  This moves the qubit
$|c_0\rangle$ from node A to node B.  $|c_0\rangle$ is used in
computation at node B, then moved back to node A via a similar
teleportation to complete the computation.  The two qubits
$|t_0\rangle$ and $|t_1\rangle$ are used as transceiver qubits, and
are reinitialized as part of the QEP subcircuit.

\begin{figure*}
\centerline{\scalebox{0.65}{\hbox{
\input{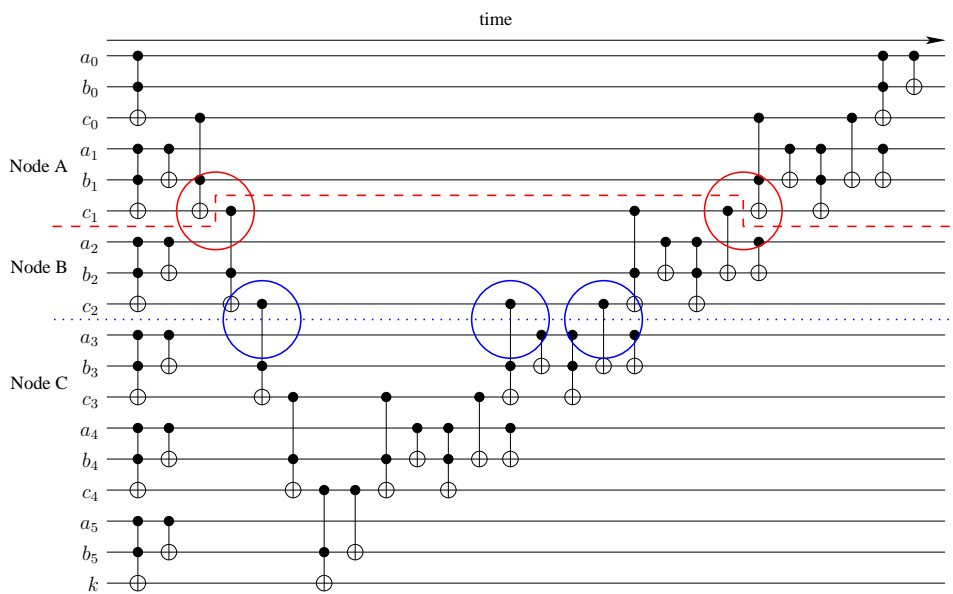}}}}
\caption{Visual approach to determining relative cost of teleporting
  data versus teleporting gates for a VBE adder.  The upper, dashed
  (red) line shows the division between two nodes (A and B) using data
  teleportation.  The circles show where the algorithm will need to
  teleport data.  The lower, dotted line (blue) shows the division
  using gate teleportation (nodes B and C).  The circles show where
  teleported gates must occur.  Note that two of these three are \textsc{ccnot}
  gates, which may entail multiple two-qubit gates in actual
  implementation.}
\label{fig:vbe-tradeoff}
\end{figure*}

Figure~\ref{fig:vbe-tradeoff} shows a larger VBE adder circuit and
illustrates a visual method for comparing telegate and teledata.  For
telegate, we can draw a line across the circuit, with the number of
gates (vertical line segments) crossed showing our cost.  For
teledata, the line must {\em not} cross gates, instead crossing the
qubit lines.  The number of such crossings is the number of
teleportations required.  This approach works well for analyzing the
VBE and CDKM adders, but care must be taken with the carry-lookahead
adder, because it uses long-distance gates that may be between
e.g. nodes 1 and 3.

The VBE adder latency to add two numbers on an $m$-node machine using
the teledata method is $2m-2$ teleportations plus the circuit cost.
For the telegate approach, using the five-gate breakdown for
\textsc{ccnot}, as in figure~\ref{fig:5gcc}, would require three
teleported two-qubit gates to form a \textsc{ccnot}.  Therefore, implementing
telegate, the latency is $7m-7$ gate teleportations, or 3.5x the cost.

\begin{figure}
\centerline{\hbox{
\input{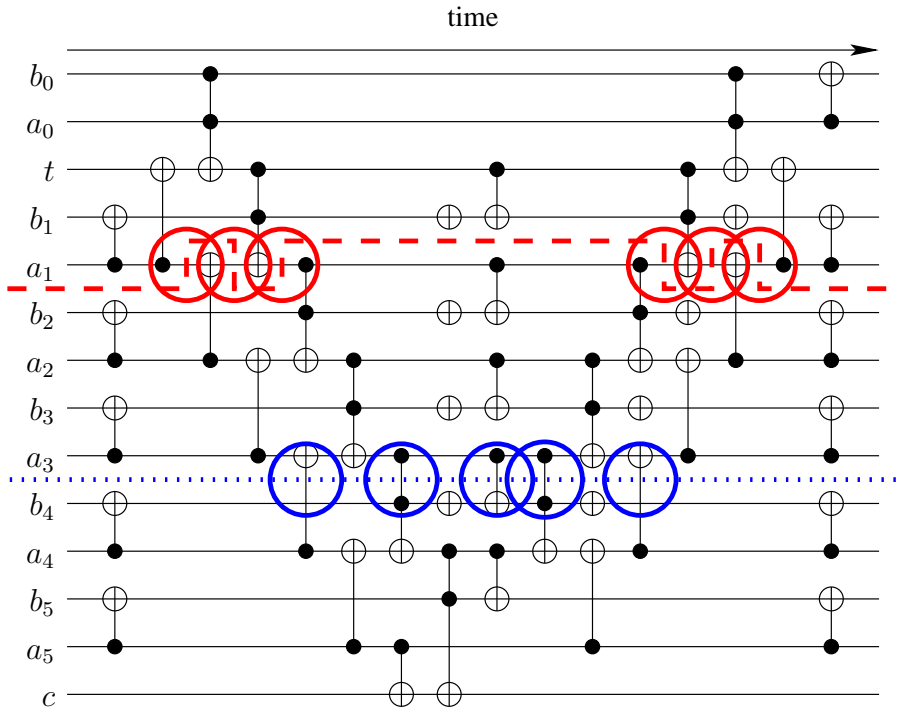}}}
\caption{Visual approach to determining relative cost of teleporting
  data versus teleporting gates for a CDKM adder.  The upper,
  dashed (red) line shows the division between two nodes using data
  teleportation.  The circles show where the algorithm will need to
  teleport data.  The lower, dotted line (blue) shows the division
  using gate teleportation.  The circles show where teleported gates
  must occur.  Note that two of these five are \textsc{ccnot} gates, which may
  entail multiple two-qubit gates in actual implementation.}
\label{fig:cuca-tradeoff}
\end{figure}

For the CDKM carry-ripple adder~\cite{cuccaro04:new-quant-ripple},
which more aggressively reuses data space, teledata requires a minimum
of six movements, whereas telegate requires two \textsc{ccnot}s and three
\textsc{cnot}s, or a total of nine two-qubit gates, as shown in
figure~\ref{fig:cuca-tradeoff}.  The CDKM adder pipelines extremely
well, so the actual latency penalty for more than two nodes is only
$2m+2$ data teleportations, or $6m$ gate teleportations, when there is
no contention for the inter-node links, as in our line and
fully-connected topologies.  The bus topology performance is limited
by contention for access to the interconnect.

\subsection{Carry-Lookahead}

Analyzing the carry-lookahead adder is more complex, as its structure
is not regular, but grows more intertwined toward the middle bits.
Gate scheduling is also variable, and the required concurrency level
is high.  The latency is $O(\log n)$, making it one of the fastest
forms of adder for large
numbers~\cite{draper04:quant-carry-lookahead,van-meter04:fast-modexp,ercegovac-lang:dig-arith}.

Let us look at the performance in a monolithic quantum computer, for
$n$ a power of two.  Based on table 1 from Draper et
al.~\cite{draper04:quant-carry-lookahead}, for $n = 2^k$, the circuit
depth of $4k+3$ Toffoli gates is 19, 31, and 43 Toffoli gates, for 16,
128, and 1,024 bits, respectively.  We assume a straightforward
mapping of the circuit to the distributed architecture.  We assign
most nodes four logical qubits ($|a_i\rangle$, $|b_i\rangle$,
$|c_i\rangle$, and one temporary qubit used as part of the carry
propagation).  In the next section, we see that the transceiver qubits
are the bottleneck; we cannot actually achieve this $4k+3$ latency.

\section{Performance}
\label{sec:perf}

The modular exponentiation to run Shor's factoring algorithm on a
1,024-bit number requires approximately 2.8 million calls to the
integer adder~\cite{van-meter04:fast-modexp}.  With a 100 $\mu$sec
adder, that will require about five minutes; with a 1 msec adder, it
will take under an hour.  Even a system two to three orders of
magnitude slower than this will have attractive performance, provided
that error correction can sustain the system state for that long, and
that the system can be built and operated economically.  This section
presents numerical estimates of adder performance which show that this
criterion is easily met by a quantum multicomputer under a variety of
assumptions about logical operation times, providing plenty of
headroom for quantum error correction.

\subsection{Initial Estimate}
\label{sec:iniperf}

Our initial results are shown in table~\ref{tab:lat-topo-baseline}.
Units are in number of complete teleportations, treating teleportation
and EPR pair generation as a single block, and assuming zero cost for
local gates.  In the following subsections these assumptions are
revisited.  We show three approaches (baseline, telegate, and
teledata) and three adder algorithms (VBE, CDKM, carry-lookahead) for
five networks (bus, 2bus, line, fully, 2fully) and three problem sizes
(16, 128, and 1024 bits).  In the baseline case, each node contains
only a single logical qubit; gates are therefore executed using the
telegate approach.  For the telegate and teledata columns, we chose
node sizes to suit the algorithms: two, three, and four qubits per
node for the CDKM, VBE, and carry-lookahead adders, respectively, when
using telegate, and three, four and five qubits when using teledata.

The VBE adder, although larger than CDKM and slower on any monolithic
computer when local gate times are considered, is faster in a
distributed environment.  The VBE adder exhibits a large (3.5x)
performance gain by using the teledata method instead of telegate.
For teledata, the performance is independent of the network topology,
because only a single operation is required at a time, moving a qubit
to a neighboring node.  The CDKM adder also communicates only with
nearest neighbors, but performs more transfers.  The single bus
configuration is almost 3x slower than the line topology.  On a line,
in most time slots, three concurrent transfers are conducted (e.g.,
between nodes $1\rightarrow 2$, $3\rightarrow 2$, and $3\rightarrow
4$).

An unanticipated but intuitive result is that the performance of the
carry-lookahead adder is better in the baseline case than the telegate
case, for the fully-connected network.  This is due to the limitation
of having a single transceiver qubit per node.  Putting more qubits in
a node increases contention for the transceiver qubit, and reduces
performance even though the absolute number of gates that must be
executed via teleportation has been reduced.  The carry-lookahead
adder is easily seen to be inappropriate for the line architecture,
since the carry-lookahead requires long-distance gates to propagate
carry information quickly.  Our numbers also show that the
carry-lookahead adder is not a good match for a bus architecture,
despite the favorable long-distance transport, again because of
excessive contention for the bus.

For telegate, performing some adjustments to eliminate intra-node
gates, we find $8n - 9k - 8$ total Toffoli gates that need arguments
that are originally stored on three separate nodes, plus $n-2$
two-node \textsc{cnot}s.  For the bus case, which allows no concurrency, this
is our final cost.  For the fully-connected network, we find a depth
of $8k-10$ three-node \textsc{ccnot}s, 8 two-node \textsc{ccnot}s, and 1 \textsc{cnot}.  These
must be multiplied by the appropriate \textsc{ccnot} breakdown.  For the
teledata fully-connected case, each three-node Toffoli gate requires
four teleportations (in and out for each of two variables).  For the
2fully network, the latency of the three-node Toffolis is halved, but
the two-node Toffolis do not benefit, giving us a final cost of
slightly over half the fully-connected network cost.
\begin{landscape}
\begin{table*}
\centerline{\hbox{
\begin{tabular}{||r|r||r|r|r||r|r|r|r|r||r|r|r|r|r||}\hline
algo. & size & \multicolumn{3}{c||}{Baseline} &
\multicolumn{5}{c||}{Telegate} &
\multicolumn{5}{c||}{Teledata} \\
\hline
& & bus & line & fully & bus & 2bus & line & fully &2F & bus & 2bus & line & fully & 2F \\ 
\hline
VBE & 16 & 360 & 305  & 182 & 105 & 105 & 105 & 105 & 105
& 30 & 30 & 30 & 30 & 30 \\
& 128 & 3048 & 2545  & 1526 & 889 & 889 & 889 & 889 & 889
& 254 & 254 & 254 & 254 & 254 \\
& 1024 & 24552 & 20465 & 12278 & 7161 & 7161 & 7161 &
7161 & 7161 & 2046 & 2046 & 2046 & 2046 & 2046 \\
\hline
CDKM & 16 & 232 & 160   & 160 & 138 & 96 & 96 & 97 & 96
& 90 & 60 & 34 & 90 & 34 \\
& 128 & 1912 & 1280 & 1280 & 1146 & 768 & 768 & 768 & 768
& 762 & 508 & 258 & 762 & 258 \\
& 1024 & 15352 & 10240 & 10240 & 9210 & 6144 & 6144 &
6145 & 6144 & 6138 & 4092 & 2050 & 6138 & 2050 \\
\hline
Carry- & 16 & 644 & N/A & 99 & 444 & 222 & N/A & 136 & 135
& 260 & 178 & N/A & 96 & 56 \\
look- & 128 & 6557 & N/A & 159 & 4901 & 2451 & N/A & 256 & 255
& 3176 & 2028 & N/A & 192 & 104 \\
ahead & 1024 & 54806 & N/A & 219 & 41502 & 20751 & N/A & 376 & 375
& 27260 & 17206 & N/A & 288 & 152 \\
\hline
\end{tabular}
}}
\caption{Estimate of latency necessary to execute various adder
  circuits on different topologies of quantum multicomputer, assuming
  monolithic teleportation blocks (Sec.~\ref{sec:iniperf}). Units are
  in number of teleportation blocks, including EPR pair creation (bus
  transaction), local gates and classical communication.  Size, length
  of the numbers to be added, in bits.  Lower numbers are faster
  (better).  '2F' denotes 2fully.}
\label{tab:lat-topo-baseline}
\end{table*}
\end{landscape}

\subsection{Improved Performance}
\label{sec:pipeperf}

The analysis in section~\ref{sec:iniperf} assumed that a teleportation
operation is a monolithic unit.  However,
figure~\ref{fig:teleport-2bit-adder} makes it clear that a
teleportation actually consists of several phases.  The first portion
is the creation of the entangled EPR pair.  The second portion is
local computation and measurement at the sending node, followed by
classical communication between nodes, then local operations at the
receiving node.  The EPR pair creation is not data-dependent; in can
be done in advance, as resources (bus time slots, qubits) become
available, for both telegate and teledata.

Our initial execution time model treats local gates and classical
communication as free, assuming that EPR pair creation is the most
expensive portion of the computation.  For example, for the teledata
VBE adder on a linear topology, all of the EPR pairs needed can be
created in two time steps at the beginning of the computation.  The
execution time would therefore be 2, constant for all $n$.
Table~\ref{tab:lat-topo-pipeline} shows the performance under this
assumption.  The performance of the carry-lookahead adder does not
change, as the bottleneck link is busy full-time creating EPR pairs.

This model gives a misleading picture of performance once EPR pair
creation is decoupled from the teleportation sequence.  When the cost
of the teleportation itself or of local gates exceeds $\sim 1/n$ of
the cost of the EPR pair generation, the simplistic model breaks down;
in the next subsection, we examine the performance with a more
realistic model.
\begin{landscape}
\begin{table*}
\centerline{\hbox{
\begin{tabular}{||r|r||r|r|r||r|r|r|r|r||r|r|r|r|r||}\hline
algo. & size & \multicolumn{3}{c||}{Baseline} &
\multicolumn{5}{c||}{Telegate} &
\multicolumn{5}{c||}{Teledata} \\
\hline
& & bus & line & fully & bus & 2bus & line & fully &2F & bus & 2bus & line & fully & 2F \\ 
\hline
VBE & 16 & 360 & 16  & 16 & 105 & 53 & 7 & 14 & 7
& 30 & 15 & 2 & 4 & 2 \\
& 128 & 3048 & 16  & 16 & 889 & 445 & 7 & 14 & 7
& 254 & 127 & 2 & 4 & 2 \\
& 1024 & 24552 & 16 & 16 & 7161 & 3581 & 7 &
14 & 7 & 2046 & 1023 & 2 & 4 & 2 \\
\hline
CDKM & 16 & 232 & 21   & 19 & 135 & 68 & 11 & 18 & 9
& 90 & 60 & 6 & 12 & 6 \\
& 128 & 1912 & 21 & 19 & 1146 & 573 & 11 & 18 & 9
& 762 & 508 & 6 & 12 & 6 \\
& 1024 & 15352 & 21 & 19 & 9210 & 4605 & 11 &
18 & 9 & 6138 & 4092 & 6 & 12 & 6 \\
\hline
Carry- & 16 & 644 & N/A & 99 & 444 & 222 & N/A & 89 & 45
& 260 & 178 & N/A & 96 & 56 \\
look- & 128 & 6557 & N/A & 159 & 4901 & 2451 & N/A & 149 & 75
& 3176 & 2028 & N/A & 192 & 104 \\
ahead & 1024 & 54806 & N/A & 219 & 41502 & 20751 & N/A & 209 & 105
& 27260 & 17206 & N/A & 288 & 152 \\
\hline
\end{tabular}
}}
\caption{Estimated latency to execute various adders on different
  topologies, for decomposed teleportation blocks
  (sec.~\ref{sec:pipeperf}), assuming classical communication and
  local gates have zero cost.  Units are in EPR pair creation times.
  '2F' denotes 2fully.
}
\label{tab:lat-topo-pipeline}
\end{table*}
\end{landscape}

\subsection{Detailed Estimate}

\begin{figure*}
\centering
\subfigure{
\includegraphics[width=.45\textwidth]{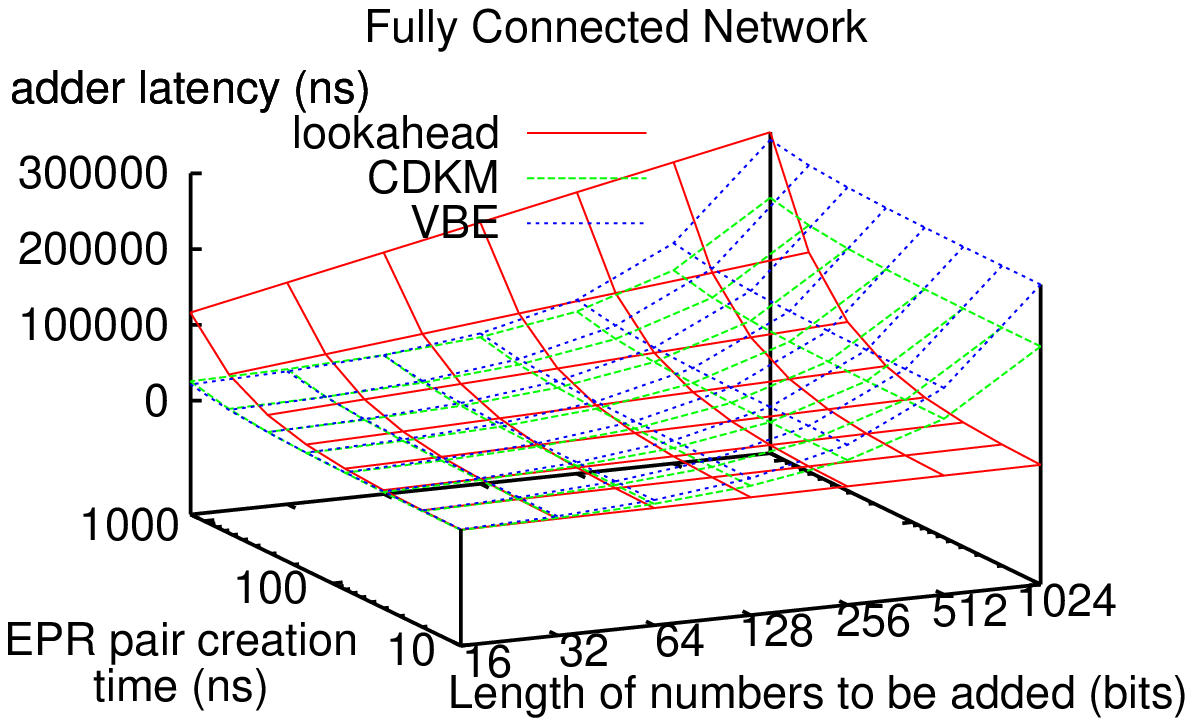}}
\hspace{.3in}
\subfigure{
\includegraphics[width=.45\textwidth]{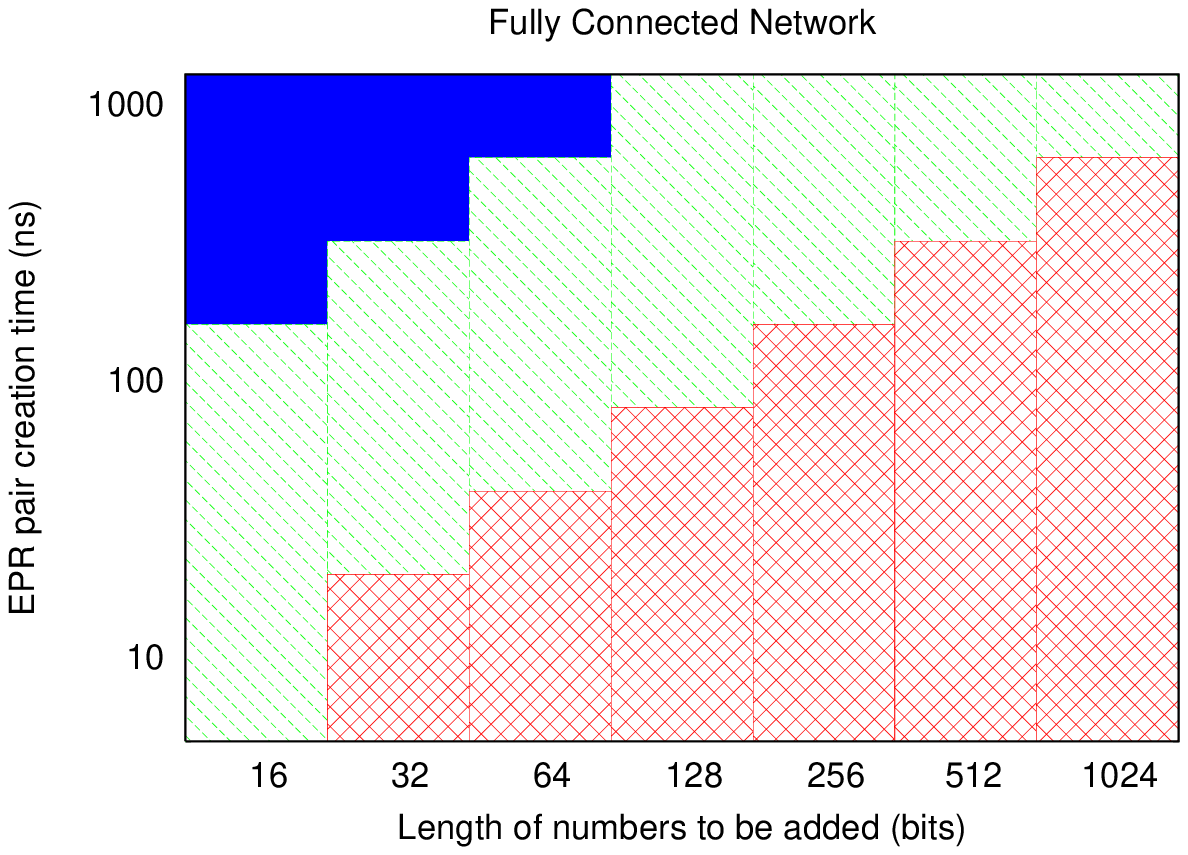}}
\hspace{.3in}
\subfigure{
\includegraphics[width=.45\textwidth]{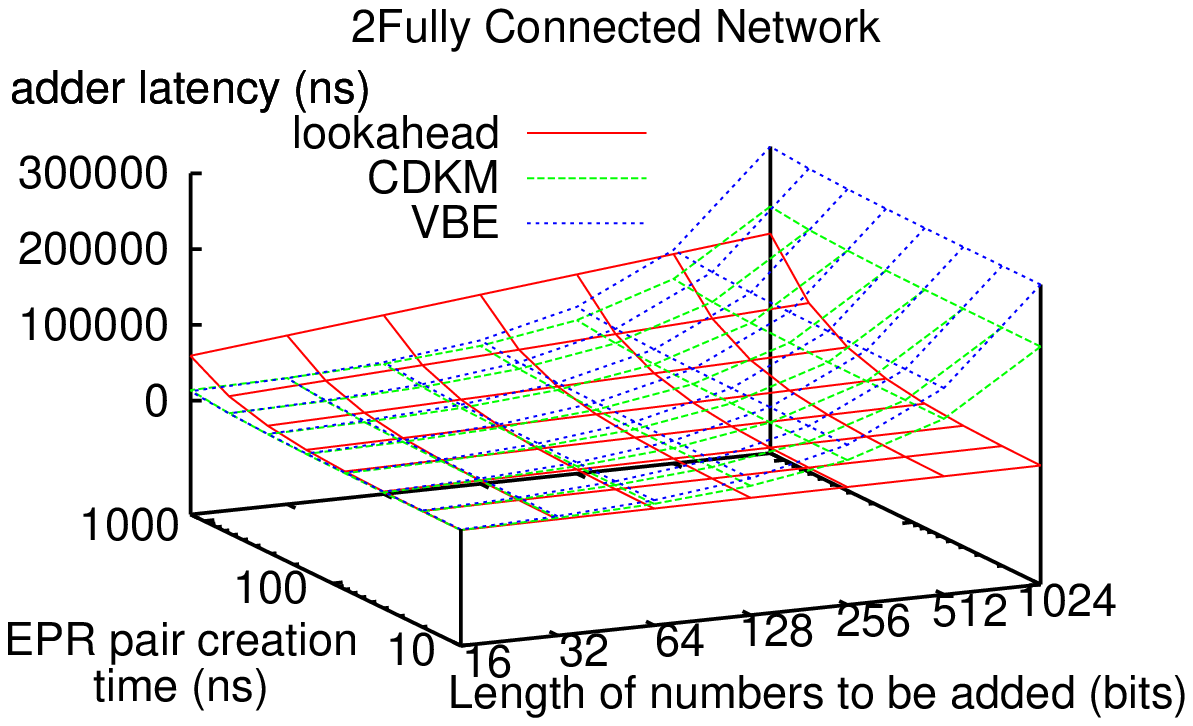}}
\hspace{.3in}
\subfigure{
\includegraphics[width=.45\textwidth]{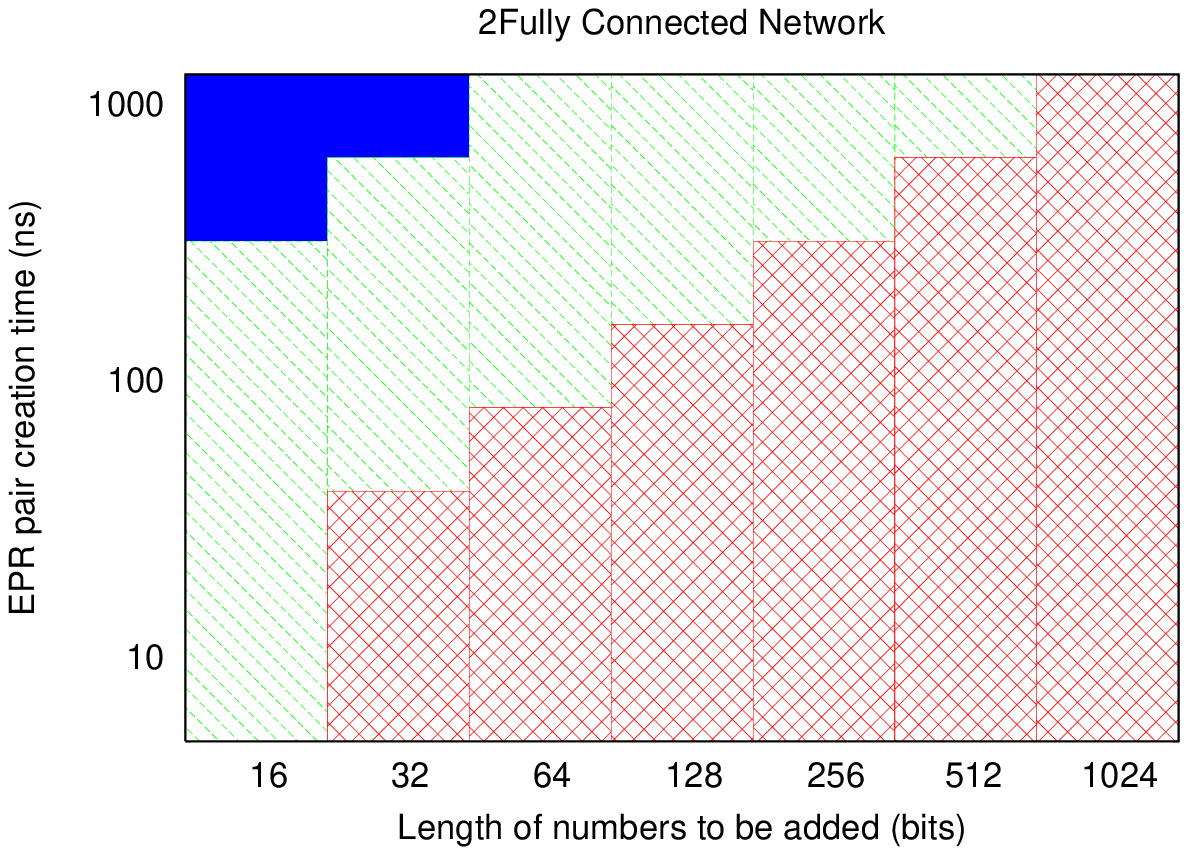}}
\hspace{.3in}
\subfigure{
\includegraphics[width=.45\textwidth]{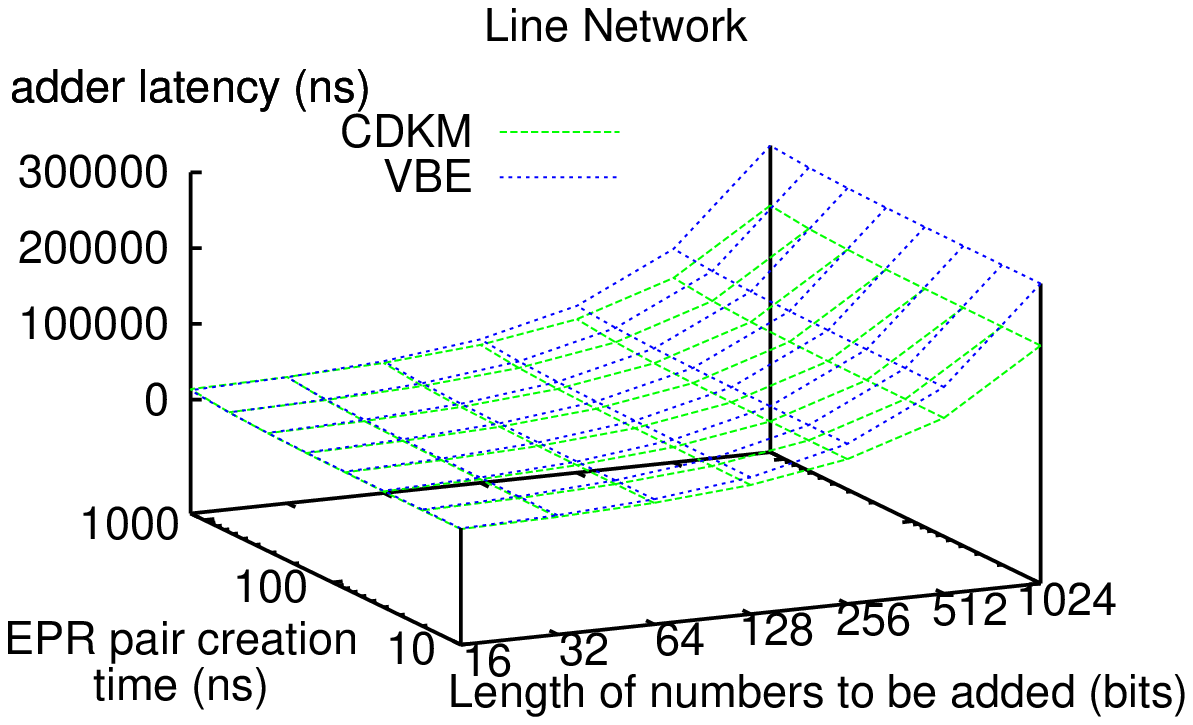}}
\hspace{.3in}
\subfigure{
\includegraphics[width=.45\textwidth]{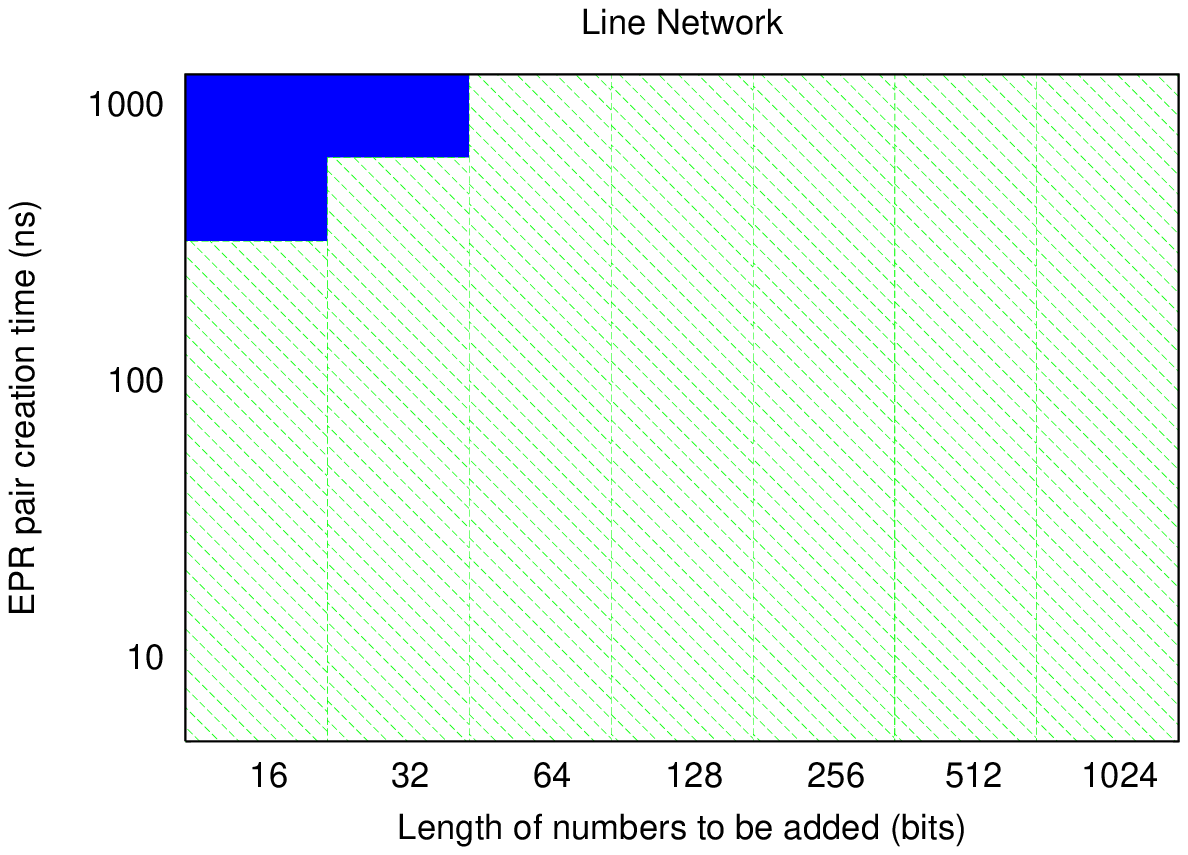}}
\caption{(Telegate) Performance of different adders on three different
  networks, one fully-connected with a single link and one with two
  links per node (2fully), and one line configuration.  In this graph,
  we vary the latency to create a high-quality EPR pair and the length
  of the numbers we are adding.  Classical communication time is
  assumed to be 10nsec, Toffoli gate time 50nsec, \textsc{cnot} gate time
  10nsec.  The left hand graph of each pair plots adder execution time
  (vertical axis) against EPR pair creation time and number length.
  In the right hand graph of each pair, the hatched red area indicates
  areas where carry-lookahead is the fastest, the diagonally lined
  green area indicates CDKM carry-ripple, and solid blue indicates VBE
  carry-ripple.  The performance of the carry-lookahead adder is very
  sensitive to the EPR pair creation time.  If EPR pair creation time
  is low, the carry-lookahead adder is very fast; if creation time is
  high, the adder is very slow.}
\label{fig:fully}
\end{figure*}

\begin{figure*}
\centering
\subfigure{
\includegraphics[width=.45\textwidth]{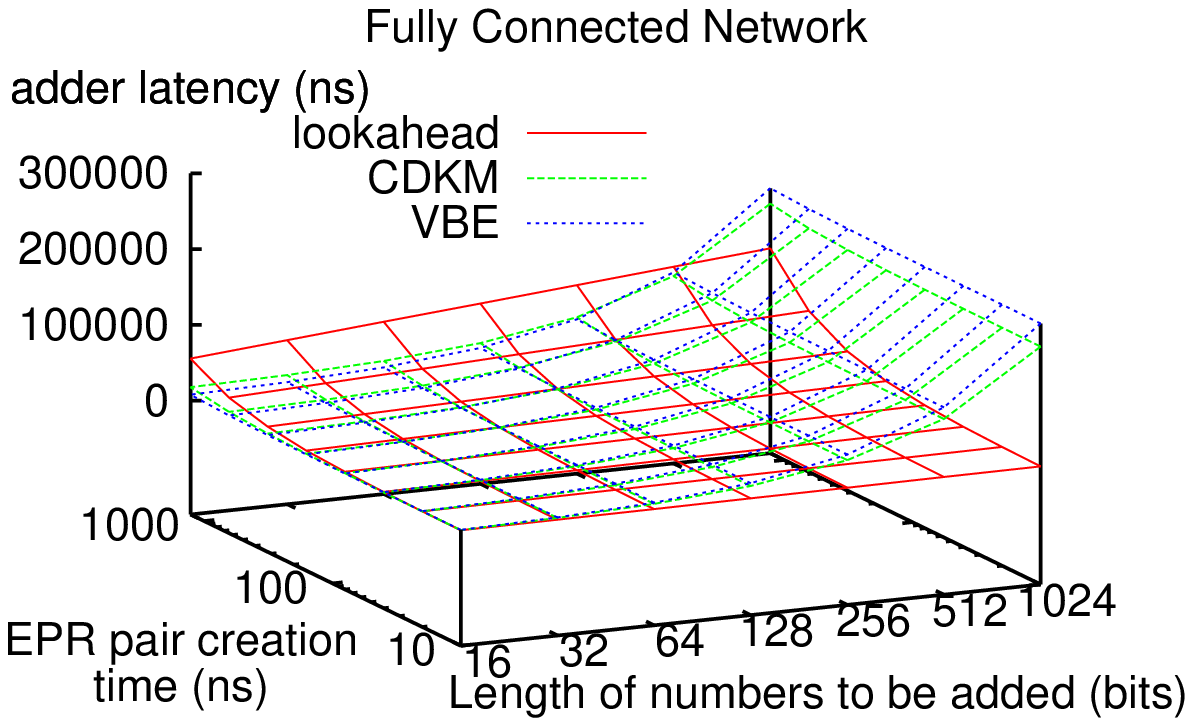}}
\hspace{.3in}
\subfigure{
\includegraphics[width=.45\textwidth]{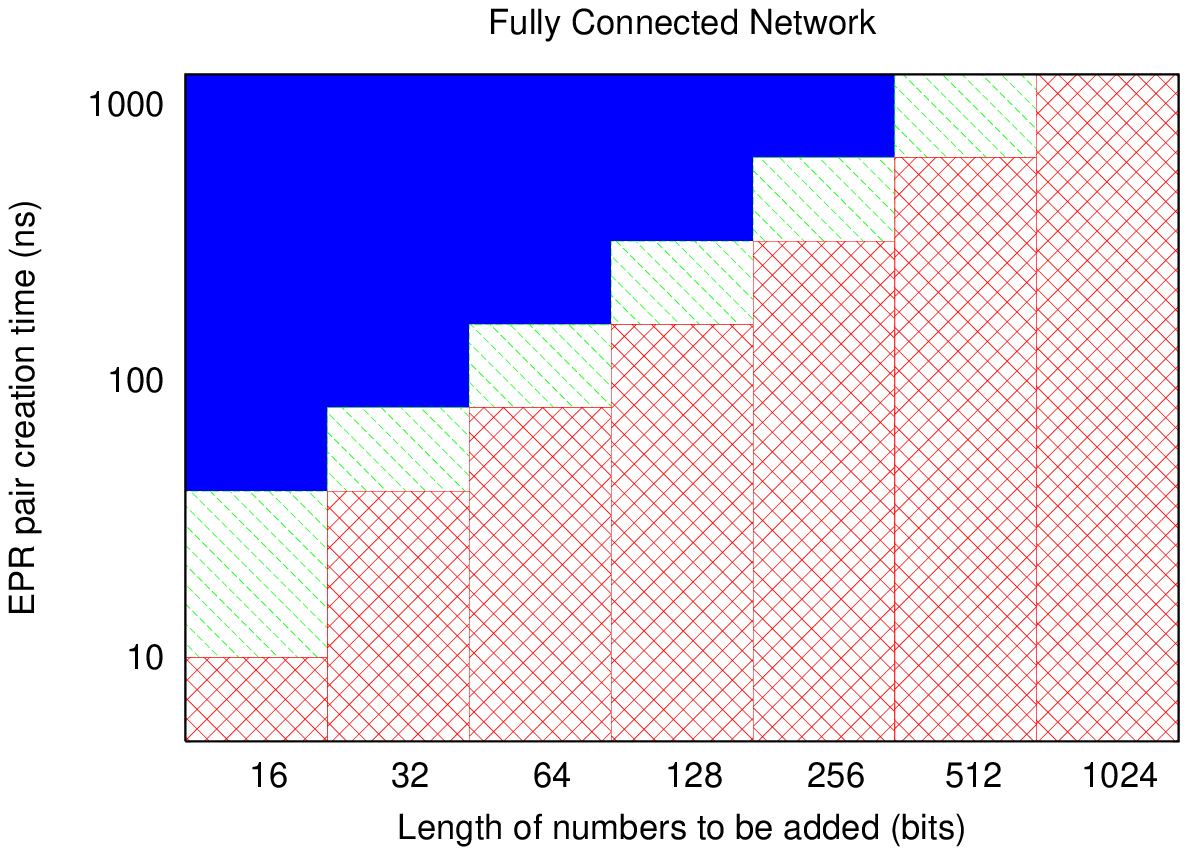}}
\hspace{.3in}
\subfigure{
\includegraphics[width=.45\textwidth]{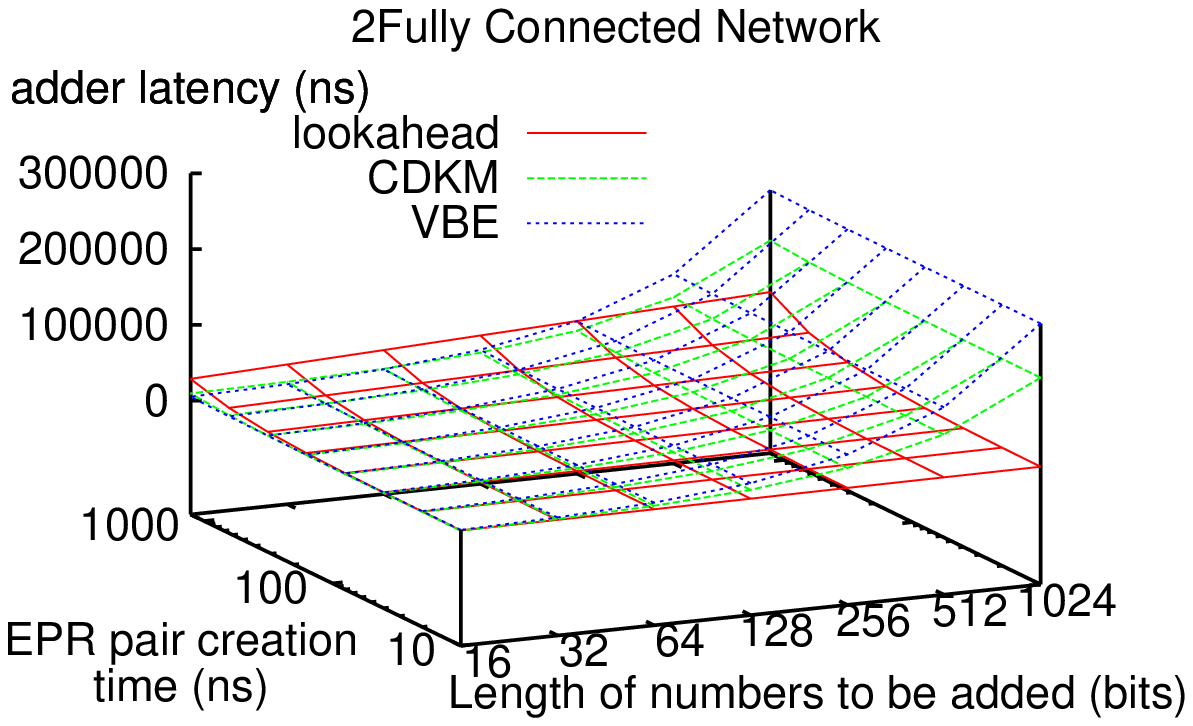}}
\hspace{.3in}
\subfigure{
\includegraphics[width=.45\textwidth]{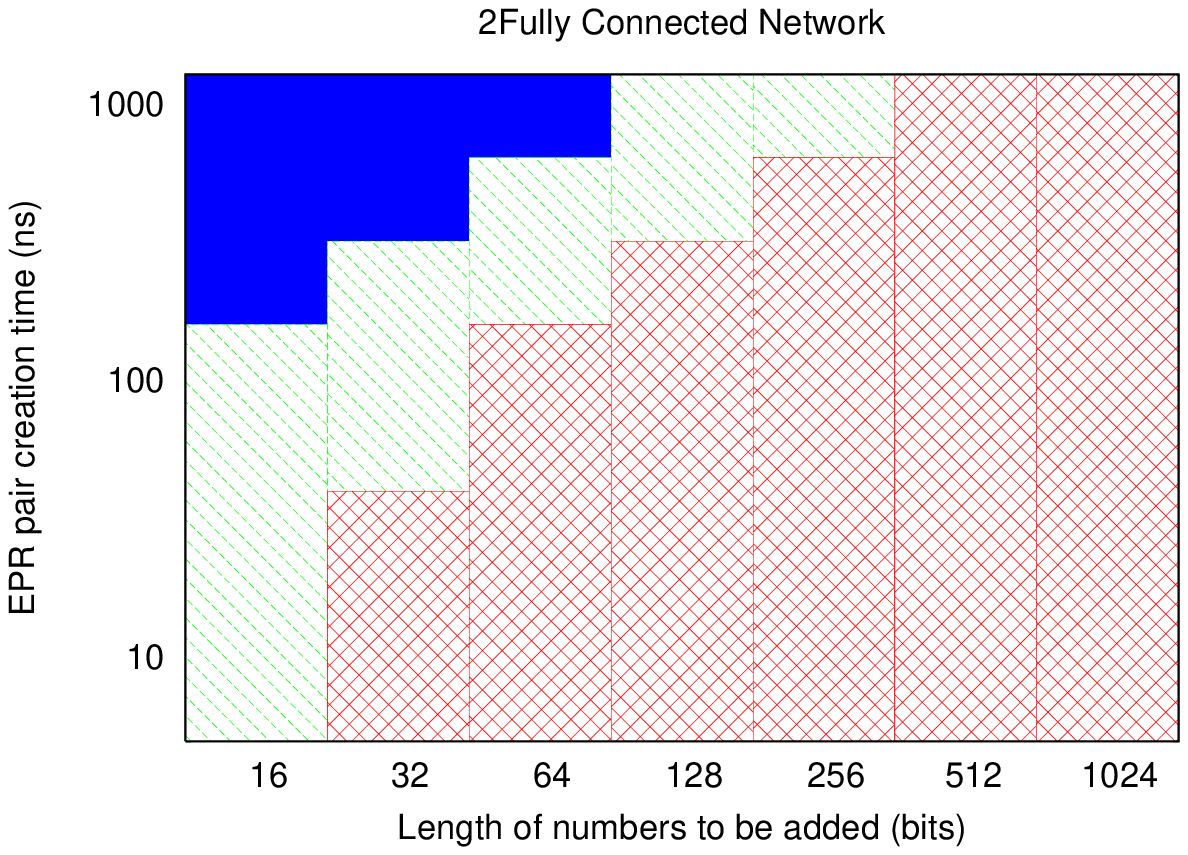}}
\hspace{.3in}
\subfigure{
\includegraphics[width=.45\textwidth]{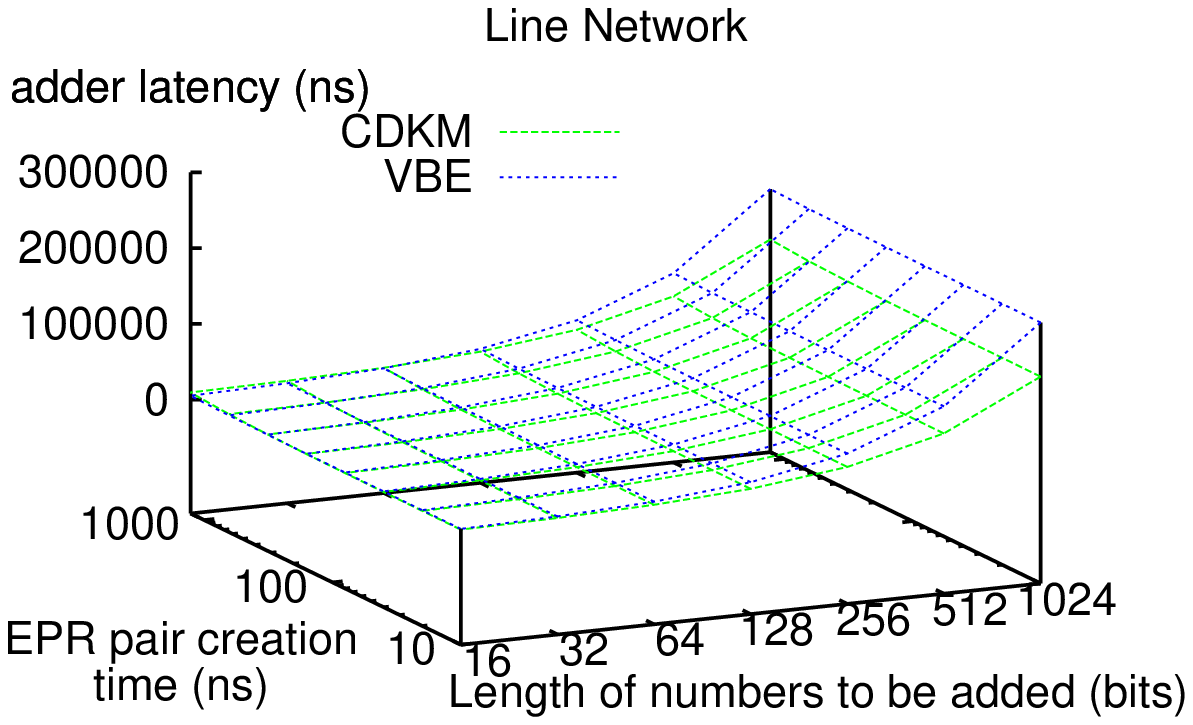}}
\hspace{.3in}
\subfigure{
\includegraphics[width=.45\textwidth]{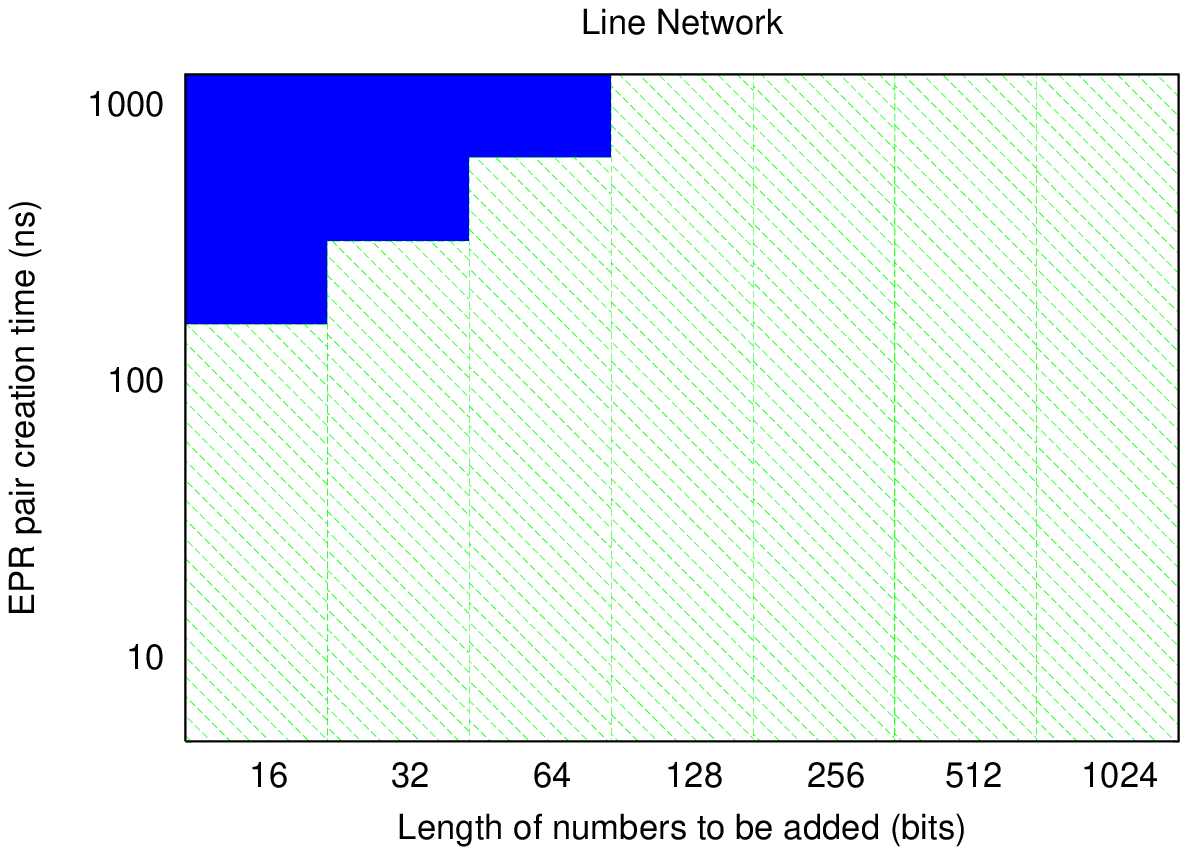}}
\caption{(Teledata) Performance of different adders on three different
  networks, one fully-connected with a single link and one with two
  links per node (2fully), and one line configuration.  In this graph,
  we vary the latency to create a high-quality EPR pair and the length
  of the numbers we are adding.  Classical communication time is
  assumed to be 10nsec, Toffoli gate time 50nsec, \textsc{cnot} gate time
  10nsec.  In the right hand graph of each pair, the hatched red area
  indicates areas where carry-lookahead is the fastest, the diagonally
  lined green indicates CDKM carry-ripple, and solid blue indicates
  VBE carry-ripple.  The performance of the carry-lookahead adder is
  very sensitive to the EPR pair creation time.  If EPR pair creation
  time is low, the carry-lookahead adder is very fast; if creation
  time is high, the adder is very slow.}
\label{fig:fully-td}
\end{figure*}

\begin{figure*}
\centering
\subfigure{
\includegraphics[width=.45\textwidth]{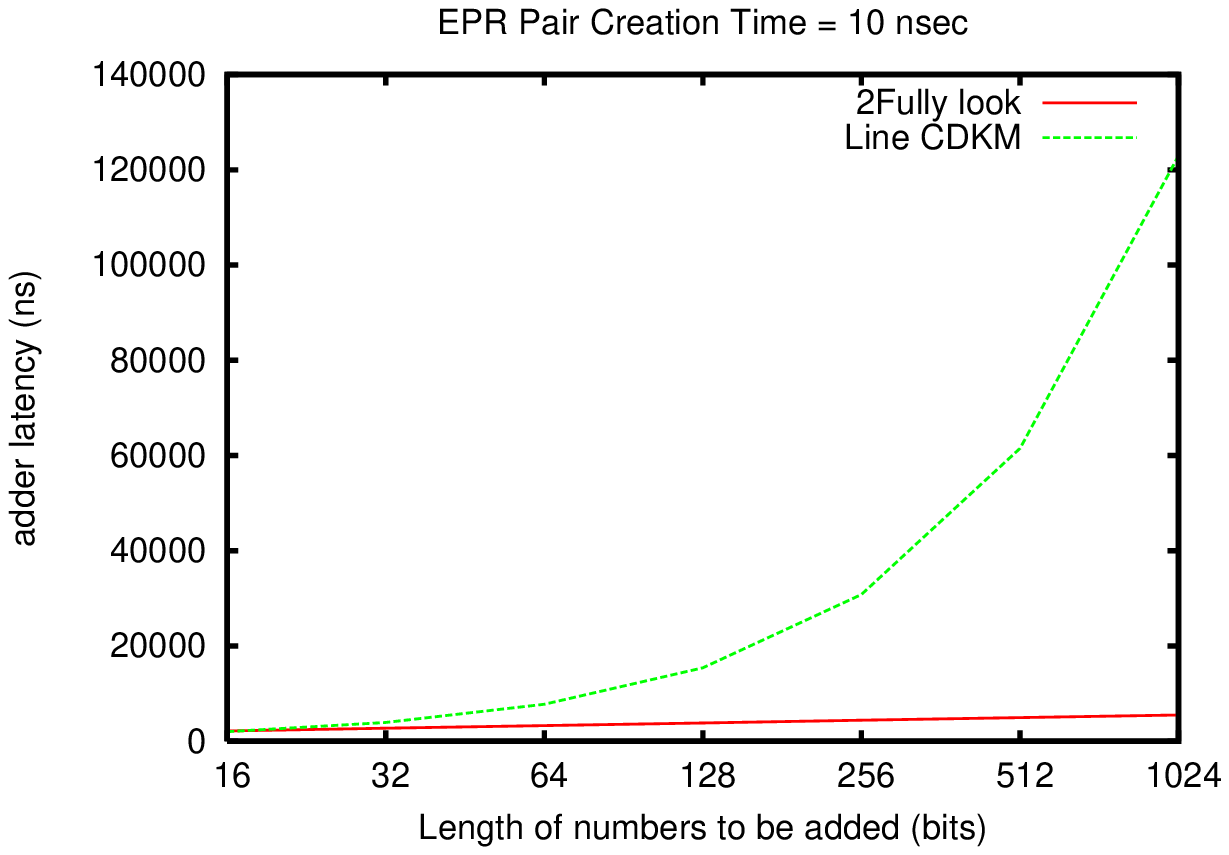}}
\hspace{.3in}
\subfigure{
\includegraphics[width=.45\textwidth]{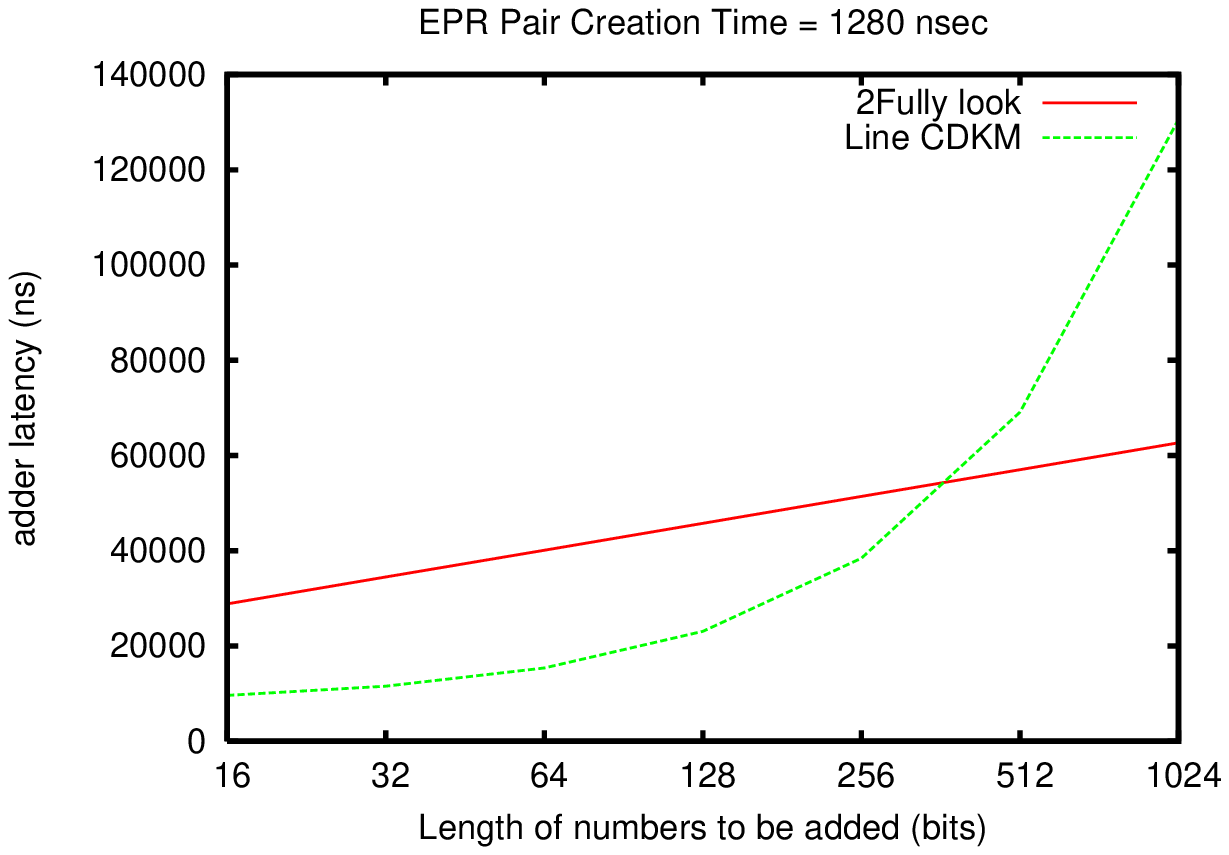}}
\caption{(Teledata) Comparison of CDKM on a line network with
  carry-lookahead on a 2fully network.  These are the ``front'' and
  ``back'' cross-sections of figure~\ref{fig:fully-td}.}
\label{fig:fully-xsec}
\end{figure*}

To create figures~\ref{fig:fully}-\ref{fig:fully-xsec}, we make
assumptions about the execution time of various operations.  Classical
communication between nodes is 10nsec.  A \textsc{ccnot} (Toffoli) gate on
encoded qubits takes 50nsec, \textsc{cnot} 10nsec, and \textsc{not} 1nsec.
These numbers can be considered realistic but optimistic for
a technology with physical gate times in the low nanoseconds; for
quantum error correction-encoded solid-state systems, the bottleneck
is likely to be the time for qubit initialization or reliable
single-shot measurement, which is still being
designed (see the references in~\cite{van-meter:qarch-impli}).

We vary the EPR pair creation time from 10nsec to 1280nsec.  This
creation process is influenced by the choice of parallel or serial bus
and the cycle time of an optical homodyne detector.  Photodetectors
may be inherently fast, but their performance is limited by
surrounding
electronics~\cite{armen02:_adaptive-homodyne,stockton02:_fpga-homodyne}.
Final performance may be faster or slower than our model, but the
range of values we have analyzed is broad enough to demonstrate
clearly the important trends.

Figures~\ref{fig:fully} and \ref{fig:fully-td} show, top to bottom,
the fully, 2fully, and line networks for the telegate and teledata
methods.  We plot adder time against EPR pair creation time and the
length of the numbers to be added.  The left hand plot shows the shape
of the surfaces, with the $z$ axis being latency to complete the
addition.  The right hand plot, with the same $x$ and $y$ axes, shows
the regions in which each type of adder is the fastest.

By examining the vertical extent of the curves in the figures, we see
that the teledata method is faster than telegate for all of the
conditions presented, but especially for the carry-lookahead adder.
The figures also show that the carry-lookahead adder is very dependent
on EPR pair creation time, while neither carry-ripple adder is. In
figure~\ref{fig:fully-xsec} we show this in more detail.  For fast
(10nsec) EPR pair creation, the carry-lookahead adder is faster for
all problem sizes.  For slow (1280nsec) EPR pair creation time,
carry-lookahead is not faster until we reach 512 bits.

Although we do not include graphs, we have also varied the time for
classical communication and the other types of gates.  The performance
of an adder is fairly insensitive to these changes; it is dominated by
the relationship between \textsc{ccnot} and EPR pair creation times.

\subsection{Comparison to a Monolithic Machine}

Our work shows the possibility of extending the size of problems that
can successfully attacked using a given quantum computing technology,
surpassing the limitations of a single, monolithic quantum computer by
aggregating many small quantum computers into one large system.
Therefore, the common multiprocessing performance analysis approach of
determining the speedup achieved by adding nodes to the system is not
appropriate.  Instead, let us ask what performance penalty we pay by
performing the computation on a multicomputer relative to the
performance of a monolithic computer, if a large enough one could be
built.

Let us examine the case of increasing node sizes connected in a linear
network, using a CDKM carry-ripple adder.  So far, we have assumed
that an $n$-bit addition is performed on a system of $n$ nodes (or
$2n$ for the baseline case).  Let us now allow the number of nodes to
be $m$, $m < n$.  If $t_{{\rm QEP}}$ is our EPR pair creation time,
$t_{{\rm CCNOT}}$ is the \textsc{ccnot} time, and $t_{{\rm TELE}}$ is our
teleportation time, including one measurement, classical communication
time between two nodes, and three local single-qubit gates (all for
logical qubits), then the total execution time on the multicomputer is
approximately
\begin{equation}
2t_{{\rm QEP}} + (m-1)t_{{\rm TELE}} + (2n-1)t_{{\rm CCNOT}}
\end{equation}
compared to $(2n-1)t_{{\rm CCNOT}}$ on a monolithic machine.  As $n$
grows and as $m\rightarrow n$, the performance penalty goes to $\sim
\frac{t_{{\rm TELE}}+2t_{{\rm CCNOT}}}{2t_{{\rm CCNOT}}}$.  For our
multicomputer environment, the cost of a logical \textsc{ccnot} is far
higher than the classical communications cost.  For small $n$, the QEP
time may dominate, but it quickly becomes an unimportant factor as $n$
grows.  Overall, the performance penalty is small, and the
capabilities improve, so we conclude that the multicomputer is a
desirable architecture.

For the carry-lookahead adder on the fully-connected and 2fully
networks, as noted above, the analysis is more difficult because the
transceiver qubits are often the bottleneck in the system.  We compare
only the fixed node size of five logical qubits and the 2fully network
to a monolithic machine of approximately $4n$ logical qubits.  In this
configuration, the performance penalty is significant; $2\times$ for
an EPR pair creation time of 10nsec, and $25\times$ for an EPR pair
creation time of 1280nsec, almost independent of $n$, as both the
computational circuit depth and the communication overhead scale with
$O(\log n)$.  Of course, the carry-lookahead adder in general is
favorable for very large $n$, so despite the apparently large
performance penalty it may still be the preferred choice in that case.

\section{Conclusion}

A quantum multicomputer is a system composed of multiple nodes, each
of which is a small quantum computer capable of creating entanglement
shared with other nodes via a qubus.  We have evaluated the
performance of arithmetic circuits on a quantum multicomputer for
different problem sizes, interconnect topologies, and gate timings.
Although we have assumed that the interconnect is based on the qubus
entanglement protocol creation of EPR pairs, our analysis, especially
table~\ref{tab:lat-topo-baseline}, applies equally well to any
two-level structure with low-latency local operations and high-latency
long-distance operations.  The details of the cost depend on the
interconnect topology, number of transceiver qubits, and the chosen
breakdown for \textsc{ccnot}.  More important than actual gate times
for this analysis is gate time ratios.  The time values presented here
are reasonable for solid-state qubits under optimistic assumptions
about advances in the underlying technology.  Applying our results to
slower technologies (or the same technology using more layers of
quantum error correction) is a simple matter of scaling by the
appropriate clock speed and storage requirements.

We find that the teledata method is faster than the telegate method,
that separating the actual data teleportation from the necessary EPR
pair creation allows a carry-ripple adder to be efficient for large
problems, and that a linear network topology is adequate for up to a
hundred nodes or more, depending on the cost ratio of EPR pair
creation to local gates.  For very large systems, switching
interconnects, which are well understood in the optical
domain~\cite{kim03:_1100_port_mems,marchand:opt-aug-3d-computer,szymanski:_design-terabit-photonic-backplane},
may become necessary, though we recommend deferring adding switching
due to the complexity and the inherent signal loss; switching time in
such systems also must be considered.

Our results show that node size, interconnect topology, distributed
gate approach (teledata v. telegate), and choice of adder affect
overall performance in sometimes unexpected ways.  Increasing the
number of logical qubits per node, for example, reduces the total
number of interconnect transfers but concentrates them in fewer
places, causing contention for access for some algorithms.  For the
specific recommendation of a linear network and a carry-ripple adder,
larger nodes exact no penalty but produce no benefit.  Therefore,
increasing node size is not, in general, favorable {\em unless node
I/O bandwidth increases proportionally}; we recommend keeping the node
size small and fixed for the foreseeable future.

Our data presents a clear path forward.  We recommend pursuing a node
architecture consisting of only a few logical qubits and initially two
transceiver (quantum I/O) qubits.  This will allow construction of a
linear network, which will perform adequately with a carry-ripple
adder up to moderately large systems.  Engineering emphasis should be
placed on supporting more transceiver qubits in each node, which can
be used to parallelize transfers, decrease the network diameter, and
provide fault tolerance.  Significant effort is warranted on
minimizing the key parameter of EPR pair creation time.  Only once
these avenues have been exhausted should the node size be increased
and a switched optical network introduced.  This approach should lead
to the design of a viable quantum multicomputer.

\section*{Acknowledgments}

The authors thanks Eisuke Abe for useful discussions, and Thaddeus
Ladd for both discussions and writing advice.  Professors Hideharu
Amano, Nobuyuki Yamasaki and Timothy Pinkston provided advice on
multicomputer networks.  RDV and KMI acknowledge funding from
CREST-JST.  KN acknowledges funding from MEXT.  WJM acknowledges
funding from the European project QAP.

\bibliography{paper-reviews}


\end{document}